\documentclass{amsart}
\usepackage[foot]{amsaddr}
\usepackage{lmodern}
\usepackage{amsmath}
\usepackage{mathtools}
\usepackage{amsfonts}
\usepackage{amssymb}
\usepackage{xcolor}
\usepackage{amsxtra}
\usepackage[utf8]{inputenc}
\usepackage[T1]{fontenc}
\usepackage{amsthm}
\usepackage{float}
\usepackage[pass]{geometry}
\usepackage{graphicx}
\usepackage[hidelinks]{hyperref}
\usepackage{url}
\usepackage{array}
\usepackage{orcidlink}
\usepackage[normalem]{ulem}

\usepackage[backend=biber,style=numeric-comp,doi=true,url=false,maxnames=20,sorting=none]{biblatex}

\DeclareFieldFormat[article, inproceedings]{title}{#1}
\DeclareFieldFormat[article]{journaltitle}{\mkbibemph{#1}}
\DeclareFieldFormat[inproceedings]{booktitle}{\mkbibemph{#1}}
\DeclareFieldFormat{volume}{\textbf{#1}}
\DeclareFieldFormat{pages}{#1}
\DeclareDelimFormat{finalnamedelim}{%
  \addspace
  \bibstring{and}\space
}

\DeclareBibliographyDriver{article}{%
  \printnames{author}
  \addcomma\space
  \printfield[title]{title}
  \addcomma\space
  \printfield[journaltitle]{journaltitle}
  \addcomma\space
  \textbf{\printfield{volume}}
  \addspace
  \mkbibparens{\printfield{year}}
  \iffieldundef{pages}{}{
    \addcomma\space
    \printfield{pages}
  }
  \adddot
  \adddot\par
  \iffieldundef{doi}{}{
    \noindent\href{https://doi.org/\strfield{doi}}{\texttt{https://doi.org/\strfield{doi}}}
  }
}

\DeclareBibliographyDriver{inproceedings}{%
  \printnames{author}
  \addcomma\space
  \printfield[title]{title}
  \addcomma\space
  \printfield[booktitle]{booktitle}
  \addcomma\space
  \printfield{year}
  \adddot\par
  \iffieldundef{doi}{}{
    \noindent\href{https://doi.org/\strfield{doi}}{\texttt{https://doi.org/\strfield{doi}}}
  }
}

\DeclareBibliographyDriver{misc}{%
  \printnames{author}
  \addcomma\space
  \printfield{title}
  \addcomma\space
  \printfield{howpublished}
  \iffieldundef{year}{}{\addcomma\space\printfield{year}}
  \addcomma\space
  \printfield{note}
  \adddot\par
}

\title[Chaotic Itinerancy via Entropy and Clustering]{Analysis of the Chaotic Itinerancy Phenomenon using Entropy and Clustering}

\author{Nikodem Mierski $^{1,\ast,\orcidlink{0009-0002-0937-5776}}$}
\address{\rm This is an Author Accepted Manuscript version (AAM), or post-print, of the paper published in \textit{Physical Review E} \textbf{112}, 054223 on November 24, 2025. \url{https://doi.org/10.1103/bcwd-nx7r} Copyright \copyright{} 2025 American Physical Society.}
\address{$^1$ Faculty of Applied Physics and Mathematics,
Gda\'{n}sk University of Technology,
ul.~Gabriela Narutowicza 11/12, 80-233 Gda\'{n}sk, Poland}
\email[Corresponding author]{\href{mailto:"Nikodem Mierski" <s189445@student.pg.edu.pl>}{s189445@student.pg.edu.pl}}

\address{$^\ast$ Corresponding author}

\author{Paweł Pilarczyk $^{2,\orcidlink{0000-0003-0597-697X}}$}
\address{$^2$ Faculty of Applied Physics and Mathematics \& Digital Technologies Centre,
Gda\'{n}sk University of Technology,
ul.~Gabriela Narutowicza 11/12, 80-233 Gda\'{n}sk, Poland}
\email{\href{mailto:"Paweł Pilarczyk" <pawel.pilarczyk@pg.edu.pl>}{pawel.pilarczyk@pg.edu.pl}}

\begin{document}

\begin{abstract}
We introduce a new methodology for the analysis of the phenomenon of chaotic itinerancy in a dynamical system using the notion of entropy and a clustering algorithm. We determine systems likely to experience chaotic itinerancy by means of local Shannon entropy and local permutation entropy. In such systems, we find quasi-stable states (attractor ruins) and chaotic transition states using a density-based clustering algorithm. Our approach then focuses on examining the chaotic itinerancy dynamics through the characterization of residence times within these states and chaotic transitions between them with the help of some statistical tests. We demonstrate the effectiveness of these methods on the system of globally coupled logistic maps (GCM), a well-known model exhibiting chaotic itinerancy. In particular, we conduct comprehensive computations for a large number of parameters in the GCM system and algorithmically identify itinerant dynamics observed previously by Kaneko in numerical simulations as coherent and intermittent phases.
\end{abstract}











\maketitle

\section{Introduction}
\label{sec:intro}

Chaotic itinerancy is a phenomenon observed in high-dimensional dynamical systems, often regarded as a form of intermediate behavior between order and chaos \cite{Kaneko2003-oy, Tsuda1991-pf}.
In chaotic itinerancy, trajectories are attracted to a low-dimensional ordered motion state, and stay there for a relatively long period of time. Then they depart from the ordered state and enter into high-dimensional chaotic motion. After some time, they once again reach one of ordered states, and this kind of wandering continues.

The states in which temporary stabilization occurs are called \emph{attractor ruins} because---on the one hand---they attract trajectories like a traditional attractor (an asymptotically stable set), but---on the other hand---they possess inherent instability, and thus they look like what remains from an attractor after a bifurcation. This instability often arises from the presence of unstable manifolds embedded within the quasi-attractor structure, which destabilizes trajectories despite their temporary convergence. Due to this instability, the trajectory eventually leaves the attractor ruin and transitions to another state. The order in which successive attractor ruins are visited is inherently unpredictable.
Transitions like these are often dictated by the geometry inherent to the system and are sometimes described in the literature \cite{Kaneko2003-oy, Tsuda2009-uo, Tsuda2003-dx} using the notion of Milnor attractors \cite{Milnor1985-xp} which, although not asymptotically stable, still attract a positive measure set of initial conditions.

Chaotic itinerancy can be interpreted as a specific form of metastability. This phenomenon refers to the tendency of a system to visit distinct dynamical regimes for extended periods of time before transitioning to other such regimes. Metastability can be defined in various ways, but a key aspect is the presence of long-lasting yet ultimately transient dynamical epochs \cite{Friston1997-metastability, Hancock2025-metastability, Rossi2025-metastability}. Chaotic itinerancy is thus an instance of metastability characterized by the presence of multiple, repeatedly visited regimes---attractor ruins. Many mechanisms have been proposed to explain the emergence of metastability, and describing such phenomena like chaotic itinerancy can provide insights into understanding these complex dynamical transitions.

The phenomenon of chaotic itinerancy was discovered relatively recently, and it finds applications in various practical contexts. It attracts particular attention of neuroscience, where it is applied to explain brain activity \cite{Freeman2003-ch,Freeman1987-rm, Kay2003-gl, Korn2003-pn, Miller2016-bu, Tsuda2015-ls, Tsuda2004-fq,Tsuda2004-yl}. Other applications include designing specific architectures for robotics and artificial intelligence that resemble human capabilities such as spontaneity \cite{Inoue2020-hg}. In particular, chaotic itinerancy has been proposed as a mechanism for spontaneous switching between cognitive states in working memory models, where attractor ruins correspond to distinct neural activity patterns.

Chaotic itinerancy is, in principle, easy to grasp intuitively as the alternation between ordered and chaotic dynamics. While chaotic itinerancy can be visually observed in numerical simulations of various dynamical systems, identifying the exact structure of attractor ruins and the nature of the transient chaotic states turns out to be a highly nontrivial task. The absence of a rigorous mathematical definition of chaotic itinerancy complicates efforts to rigorously analyze its properties.

This phenomenon occurs in deterministic \cite{Kaneko1993-uy, Oliveira2024,Takahisa2024-wp} and stochastic \cite{Bioni_Liberalquino2018-kt,Namikawa2005-tv} dynamical systems, as well as in neural networks \cite{Matykiewicz2004-ts, Nara1992-qg, Tsuda1992-vi, Tsuda2004-mk}. One of the simplest models in which chaotic itinerancy is observed is the system of globally coupled one-dimensional chaotic maps, such as the logistic maps \cite{Kaneko1990-fj}.

\subsection{State of the art}
\label{sec:state}

To the best of our knowledge, methods allowing one to rigorously identify and analyze chaotic itinerancy have not yet been developed. Extensive research exists, however, in which this phenomenon has been analyzed in an experimental way, mainly through numerical simulations and visualization of their results. One of the indicators that may suggest the presence of chaotic itinerancy is the slow convergence of Lyapunov exponents \cite{Sauer2003-xk, Tsuda2003-dx}. Another method of identifying parameters of a dynamical system for which chaotic itinerancy may emerge is bifurcation analysis \cite{Kobayashi2018-yv, Tanaka2005-if}. In the analysis of dynamical systems with globally coupled maps, current research focuses on the investigation of synchronization of elements. Chaotic itinerancy in these models is understood as high variability in the number of synchronized groups of elements with a given precision \cite{Kaneko1991-fh}. In spite of these efforts, no reliable methods have been developed so far that allow one to clearly determine whether chaotic itinerancy occurs in a given dynamical system or not.

\subsection{Our contribution}
\label{sec:contrib}

\begin{figure}[htbp]
\centering
\includegraphics[width=1\textwidth]{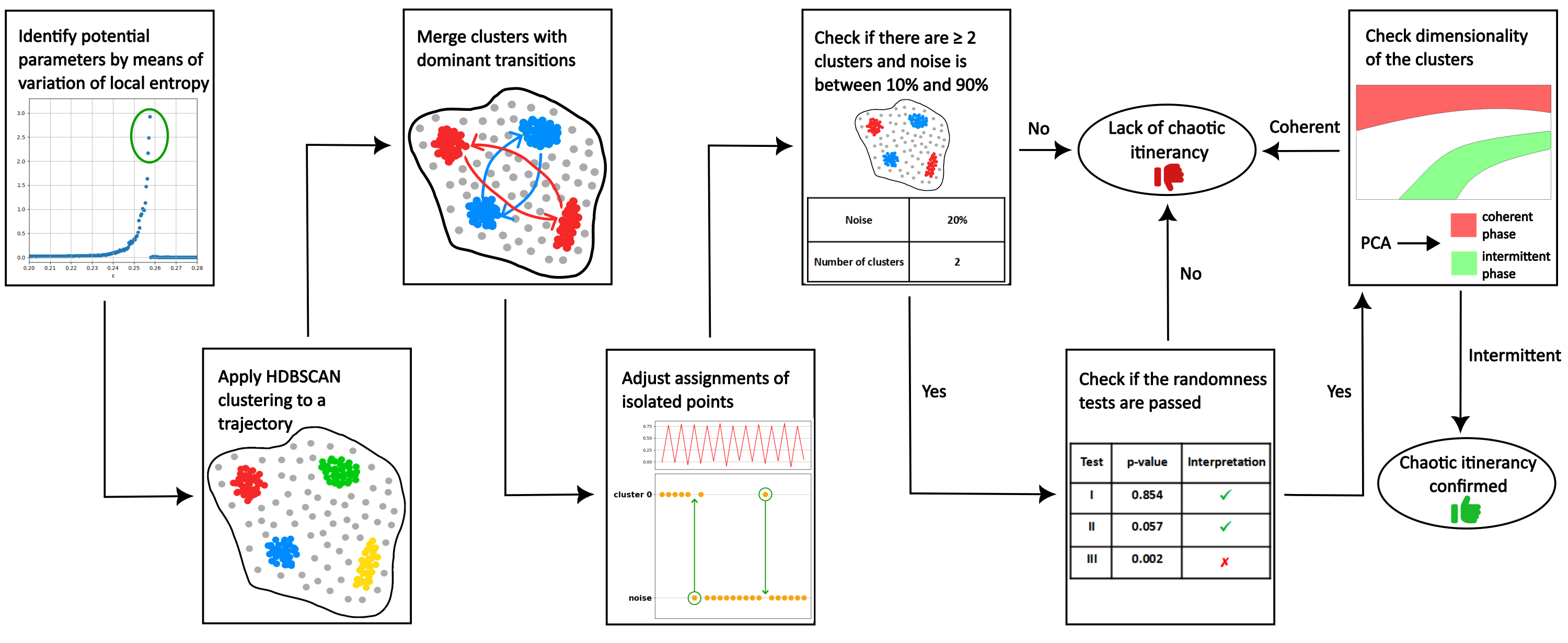}
\caption{Overview of the proposed method to determine the presence of chaotic itinerancy.}
\label{fig:overview}
\end{figure}

We introduce a method for the detection and quantification of the phenomenon of chaotic itinerancy experienced by a trajectory in a given dynamical system, using entropy to quantify the complexity of the dynamics and machine learning to find attractor ruins visited by the trajectory. Specifically, we use the hierarchical density-based clustering algorithm HDBSCAN to identify dense clusters of points that can be interpreted as attractor ruins. After assigning each point to a cluster or treating it as a transition state (``noise''), we propose a method for analyzing the characteristics of visiting the clusters by a trajectory by means of some statistical tests to exclude the possibility of ordered motion between them. Figure~\ref{fig:overview} shows an overview of this method applied to a single trajectory, with some technical details discussed in later sections. Using this approach, we develop a comprehensive method from scanning entire ranges of parameters of a dynamical system for which chaotic itinerancy may potentially be present to the analysis of the dimensionality of attractor ruins found to confirm or reject the presence of chaotic itinerancy.
We provide a software implementation of the methods introduced in this paper on~\cite{code}.

\subsection{Overview of the paper}
\label{sec:overview}

In Section \ref{sec:gcm}, we describe the system of globally coupled logistic maps (GCM), which is one of the best studied models exhibiting chaotic itinerancy. Then in Section \ref{sec:entropy}, we introduce the concept of local Shannon entropy and show how to use it to identify parameters of the GCM model that are most likely to exhibit chaotic itinerancy. In Section \ref{sec:perm}, we assess the usefulness of permutation entropy which may complement Shannon entropy in certain cases. In Section \ref{sec:dbscan}, we apply the HDBSCAN algorithm to detect dense clusters that correspond to attractor ruins. We carry out this procedure for a specific parameter point where chaotic itinerancy is expected, found by the analysis of local Shannon entropy. Once the clusters have been identified, we analyze them to characterize the attractor ruins. In Section \ref{sec:chaotic}, we investigate the dynamics in relation to the attractor ruins and provide some criteria for assessing the degree of its unpredictability (chaos). In Section \ref{sec:hdbscan}, we carry out an automated analysis of chaotic itinerancy for a wide range of parameters in the GCM model, using the HDBSCAN algorithm for clustering and PCA for determining whether the dynamics is essentially one-dimensional or more complex.

\section{Globally coupled logistic maps}
\label{sec:gcm}

Although our method for the analysis of the phenomenon of chaotic itinerancy can be applied to a variety of dynamical systems, for the sake of clarity, we shall focus on the system of globally coupled logistic maps, a system that has been widely investigated in this context.

Following Kaneko \cite{Kaneko2015-id,Kaneko2003-oy}, let us consider the $N$-dimensional dynamical system induced by the following map on the coordinates $x(i)$, with $i=1,\ldots,N$, of a point $x \in \mathbb{R}^N$:
\begin{equation}
\label{eq:gcm}
x_{n+1}(i)=(1-\varepsilon)f_a(x_n(i)) + \frac{\varepsilon}{N} \sum_{j=1}^Nf_a(x_n(j)),
\end{equation}
where $f_a \colon \mathbb{R} \ni x \mapsto 1-ax^2 \in \mathbb{R}$ is the logistic map with the parameter $a$ typically taken in a range where it commonly exhibits chaotic behavior. The parameter $a$ represents the nonlinearity of the function $f$. The parameter $\varepsilon$ takes values between 0 and 1 and determines the coupling strength between the maps. Due to the mutual dependence of the maps, the system is referred to as a system of globally coupled one-dimensional maps, or GCM for short. This model can be considered either as one map $\text{GCM}_{a,\varepsilon} \colon \mathbb{R}^N \to \mathbb{R}^N$ or as a collection of interrelated one-dimensional maps.

A characteristic property of this model, observed in numerical simulations, is the emergence of synchronization, where some elements attain nearly identical values for a long number of iterations \cite{Kaneko1990-fj, Kaneko2015-id}. Elements with nearly the same values, i.e., elements $i$ and $j$ for which $x(i) \approx x(j)$, are said to belong to the same cluster. Consequently, attractors in the system can be described by the number of clusters and the number of elements in each cluster. For different parameters of the model and different initial conditions, we observe varying numbers of clusters that are formed. Based on this, four distinct phases of the system have been identified in \cite{Kaneko1990-fj}, depending on the parameters of the system: (1) coherent phase (all elements synchronized), (2) ordered phase (few synchronized groups), (3) partially ordered phase (coexistence of configurations with many and few synchronized groups), and (4) turbulent phase (each element behaves independently). Chaotic itinerancy was then defined as the coexistence of attractors with a large number of clusters and attractors with a small number of clusters, and was observed in the partially ordered phase. However, we would like to point out that our approach introduced in this paper is different and does not rely on the relation between the individual coordinates of the iterated points.

\section{Local Shannon entropy}
\label{sec:entropy}

Local Shannon entropy is a mathematical tool that was recently proposed for testing the existence of randomness locally, as opposed to applying a global test, for example, in images \cite{Wu2013-zs}. We apply this tool to detect the possibility of chaotic itinerancy experienced by a single trajectory, represented by means of a time series.

Shannon entropy of a random variable $X$ that attains a finite number of possible values can be defined as:
\begin{equation}
H(X) = -\sum_{i=1}^{n} p_i \log_2 p_i,
\end{equation}
where $\{x_1, x_2, \ldots, x_n\}$ is the set of possible values of $X$ and $p_i = \Pr(X=x_i)$ is the probability of each value \cite{Shannon1948}.

We define \emph{local Shannon entropy} for a given point in the sequence $X = (x_i)_{i=1}^n$ as follows:
\begin{equation}
H_{\text{local}}(j) = H(X_{j-L,j+L}),
\end{equation}
where $X_{j-L,j+L}$ denotes the fragment of the sequence $X$ that includes $2L + 1$ consecutive elements from $x_{j-L}$ to $x_{j+L}$, inclusive. 
For higher-dimensional systems, we compute the sum of the values of local entropy computed for each coordinate.
Note that $H_{\text{local}}(j)$ is only defined when $L<j \leq n-L$.

\begin{figure}[tbp]
\centering
\includegraphics[width=1\textwidth]{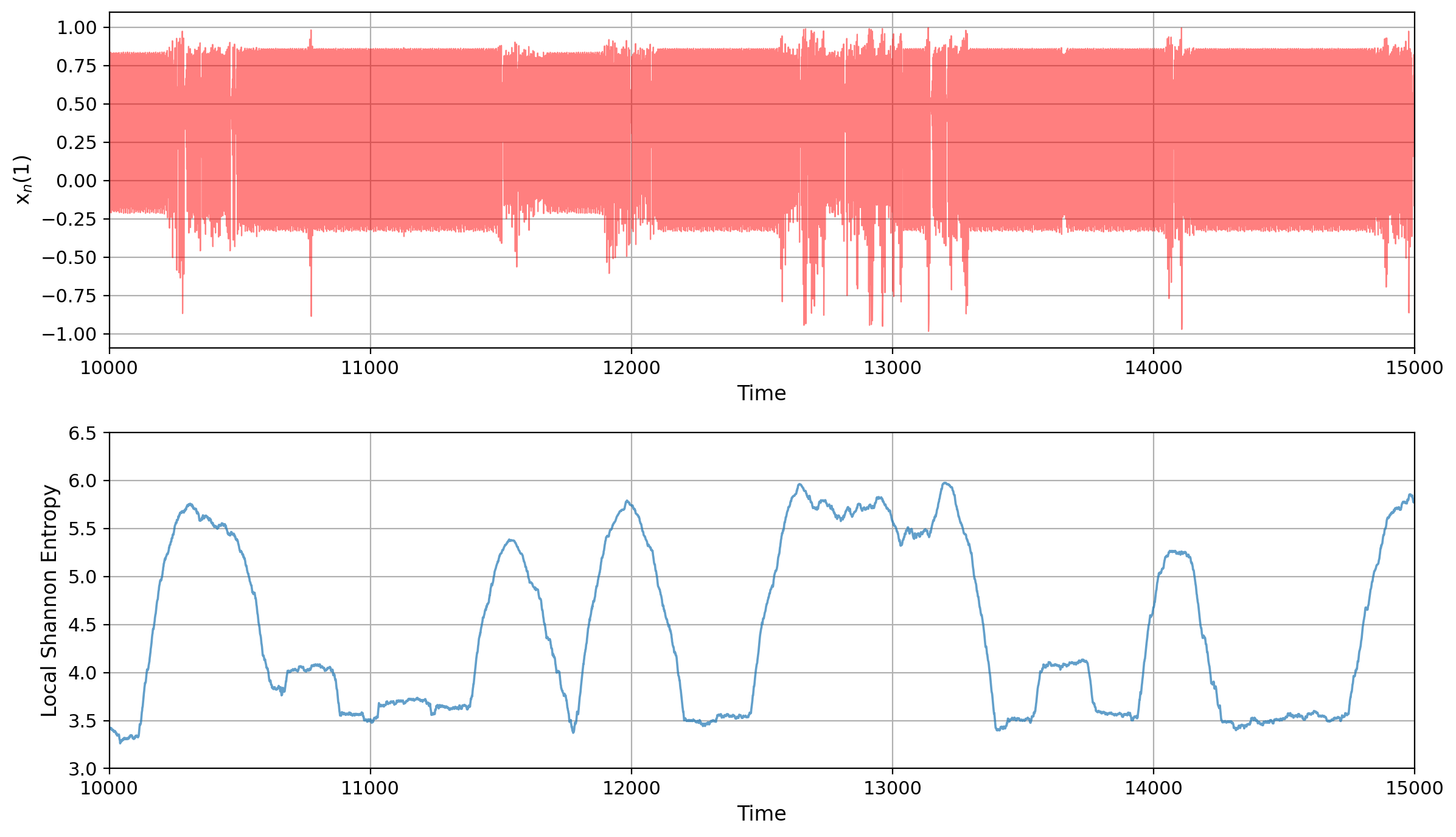}
\caption{Time series and local Shannon entropy computed for a segment of a sample trajectory for the GCM model with $N=5$, $a=2$ and $\varepsilon=0.234$.}
\label{entropy_CI}
\end{figure}

In order to choose a suitable value of $L$ it is necessary to know approximate amounts of time a trajectory typically spends wandering chaotically between attractor ruins, as well as the lengths of intervals of time when it stays in the proximity of the attractor ruins.
For this purpose, we propose to analyze the plot of a selected coordinate of the trajectory as a function of time, like the one shown in the top graph in Figure~\ref{entropy_CI}. Information on typical time intervals of ordered and transitional behavior in the system upon consideration can be read from this graph as areas of irregular fluctuations and regular changes. The radius $L$ of segments for which local entropy will be computed must be taken in such a way that the sliding window of radius $L$ can be contained in such intervals for a certain amount of time. In our case, we notice that the behavior of $x_n(1)$ is consistent in intervals of length $500$--$1000$, so we choose $L=100$ for the remainder of the paper. In fact, we also tested $L \in \{50, 200, 500\}$ and obtained almost the same results. This suggests that our method is not very sensitive to the choice of $L$.

In the case of a real-valued time series, we estimate the distribution of $X$ using a histogram. We divide the range of values into 100 bins of equal width, and we use the frequency of values falling into each bin to calculate the corresponding probability.

A system that exhibits chaotic itinerancy transitions between ordered states and a chaotic type of motion. Therefore, we expect to observe irregular fluctuations in local Shannon entropy in such a system. A high entropy value corresponds to the trajectory wandering in a high-dimensional chaotic state, while a low entropy value indicates a low-dimensional ordered state.

Figure \ref{entropy_CI} shows the values of local Shannon entropy computed for consecutive points in the case where chaotic itinerancy is observed.
One can see higher entropy values corresponding to segments of the trajectory with irregular variation of $x_n(1)$. These segments apparently correspond to chaotic wandering of the trajectory. In regions of lower entropy values, on the other hand, the graph showing $x_n(1)$ is very regular. These segments correspond to ordered motion, most likely within an attractor ruin. Note that increases in local Shannon entropy begin some time before the observed segment of irregular motion of the trajectory begins and end some time after the segment ends. Therefore, it is important to keep $L$ small enough so that temporary stabilization of the trajectory in the vicinity of an attractor ruin is not overlooked, as happens in our example around time $11900$.

\begin{figure}[tbp]
\centering
\includegraphics[width=1\textwidth]{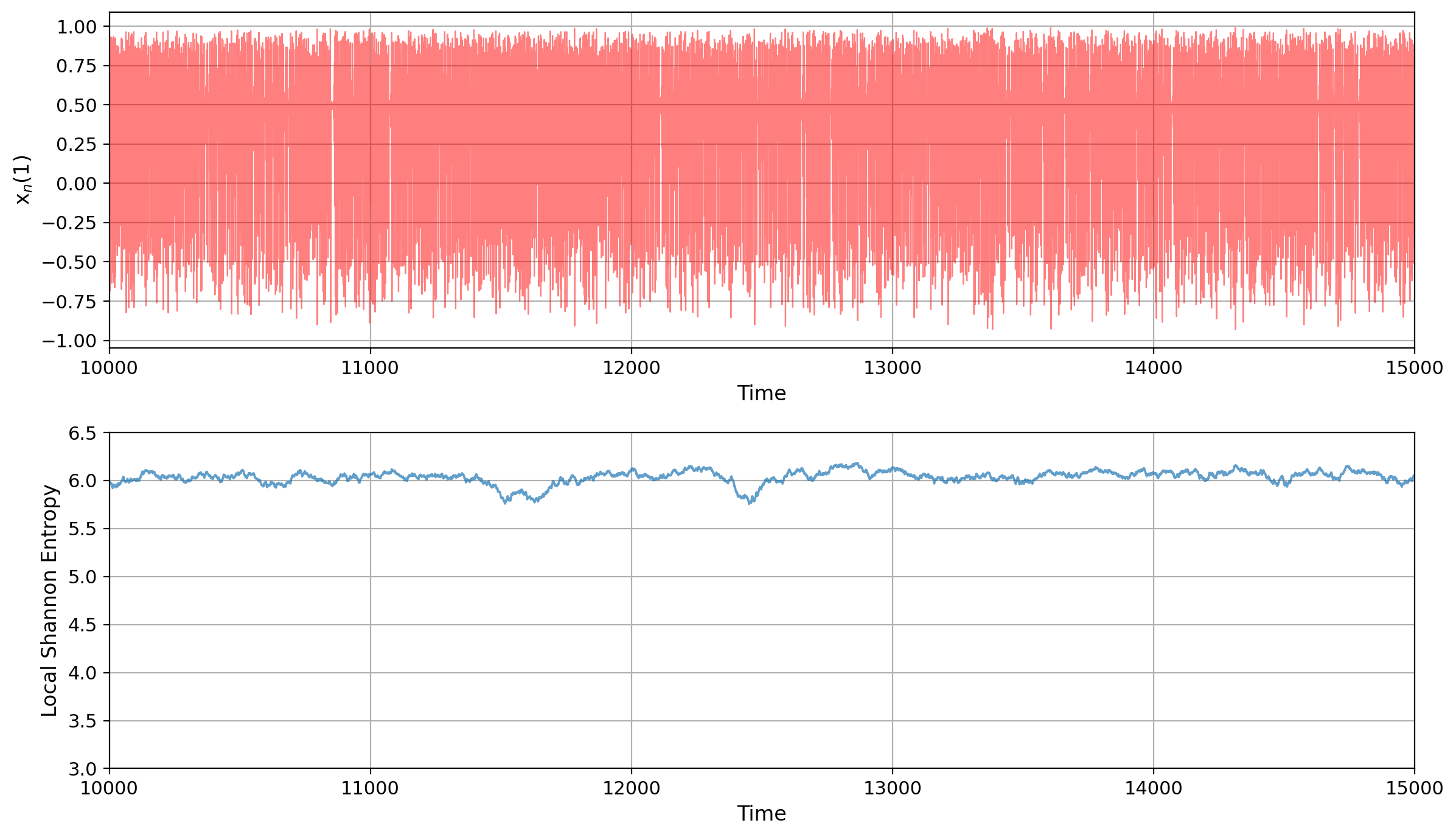}
\caption{Time series and local Shannon entropy computed for a segment of a sample trajectory for the GCM model with $N=5$, $a=2$ and $\varepsilon =0.1$.}
\label{entropy_chaotic}
\end{figure}

For comparison, Figure \ref{entropy_chaotic} shows an example in which the system exhibits purely chaotic behavior. The range of local Shannon entropy values is considerably narrower. This observation suggests that the variance of local Shannon entropy can be used to distinguish the case of chaotic itinerancy from ``classic'' chaotic dynamics. Let us discuss this next.

\begin{figure}[htbp]
\centering
\includegraphics[width=0.85\textwidth]{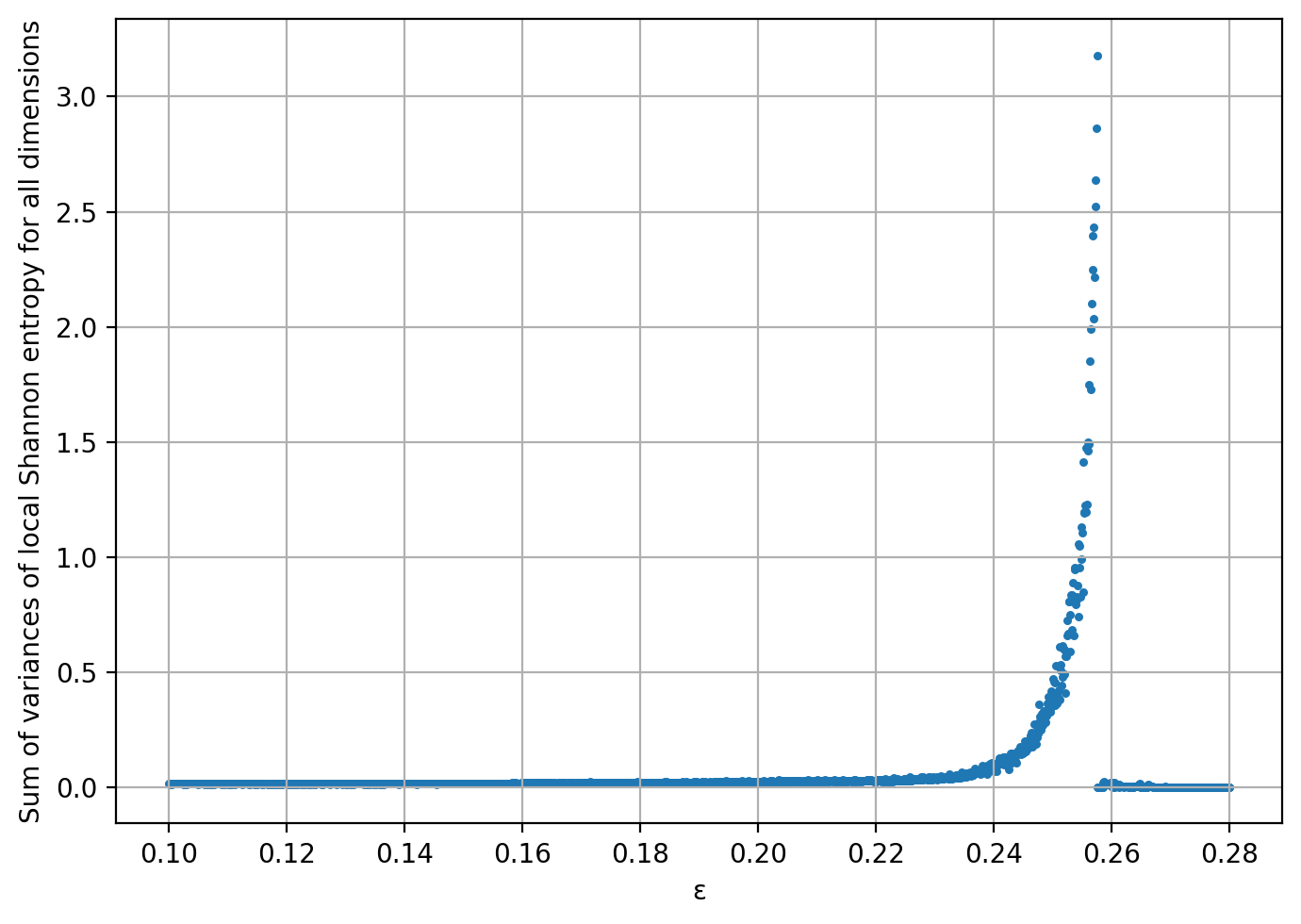}
\caption{Sum of variances of local Shannon entropy for all the coordinates of points on a segment of a sample trajectory as a function of $\varepsilon$, computed for the GCM model with $N=3$ and $a=2$.}
\label{sum_ent}
\end{figure}

Figure \ref{sum_ent} shows the variance of local Shannon entropy in the GCM model as a function of $\varepsilon$ with $N=3$ and $a=2$.
The values of $\varepsilon$ in the range from $0.1$ to $0.28$ are considered with the step of $0.0001$. For almost all values of $\varepsilon$ in this range, the variance of local Shannon entropy is close to~$0$. However, one can notice a distinct gradual increase in this variance around $\varepsilon=0.25$, which is subsequently followed by an abrupt decrease to a nearly-zero level.

\begin{figure}[htbp]
\centering
\includegraphics[width=1\textwidth]{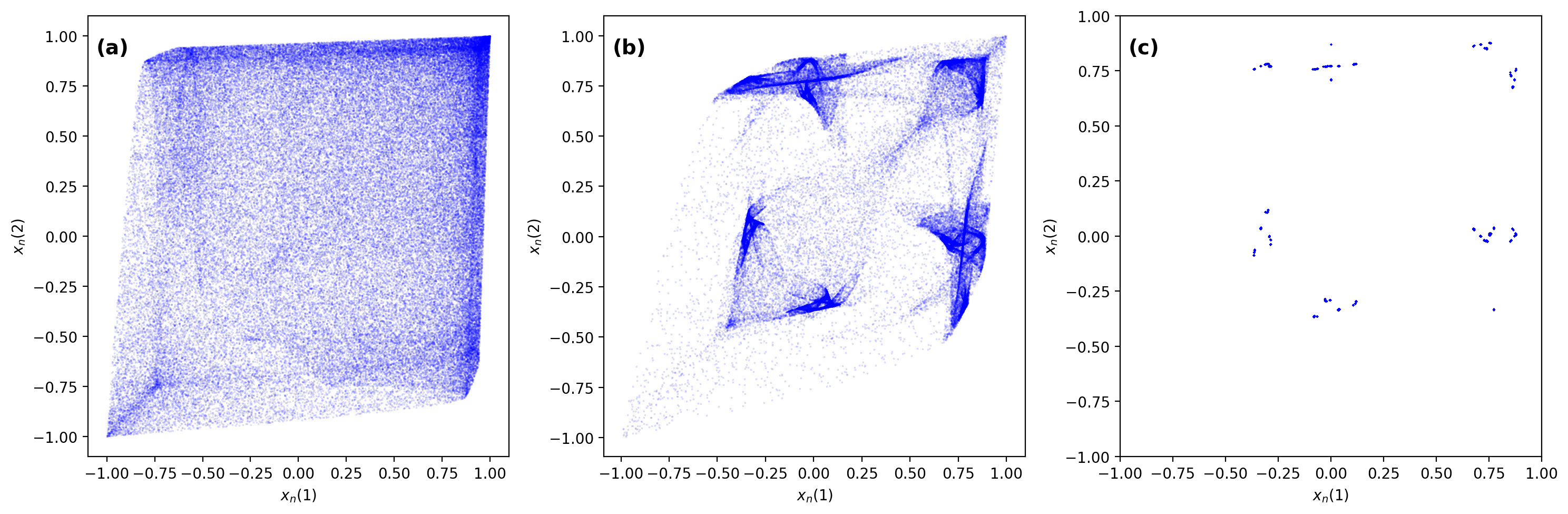}
\caption{Projection onto the first two coordinates of a sample trajectory in the GCM model with $N=3$, $a=2$ and (a) $\varepsilon=0.15$, (b) $\varepsilon=0.2574$ and (c) $\varepsilon=0.27$.}
\label{different_eps}
\end{figure}

A closer investigation of the dynamics for the different values of $\varepsilon$ reveals the following situation. A typical trajectory in the system for $\varepsilon < 0.2$ spreads in a large subset of the phase space nearly uniformly, as shown in Figure~\ref{different_eps}(a). The value of local Shannon entropy is consistently high and thus its variability shown in Figure~\ref{sum_ent} is nearly zero. As the value of $\varepsilon$ approaches $0.25$ and crosses it, some regions in the phase space emerge in which the trajectory spends considerably more time than in the remaining part of the phase space, and thus the density of points of the trajectory is clearly higher in these regions, as shown in Figure~\ref{different_eps}(b). This temporary stability is reflected in fluctuations of the local Shannon entropy and thus increased values of its variation. The dynamics complies with the idea of chaotic itinerancy, although we see this phenomenon with varying intensity, depending on the actual value of $\varepsilon$. This phenomenon is most clearly seen where the highest values of the variance of the local Shannon entropy are encountered (around $\varepsilon = 0.2574$). The high-density regions in the phase space indicate the location of attractor ruins. When $\varepsilon$ is further increased, the trajectories suddenly become attracted by one of the stable periodic orbits present in the system, starting with $\varepsilon = 0.2576$. This type of dynamics is shown in Figure~\ref{different_eps}(c). This behavior of trajectories corresponds to coherent (ordered) dynamics, reflected in low values of the local Shannon entropy. We would like to point out the fact that the values of local Shannon entropy to the left of $\varepsilon = 0.25$ and to the right of $\varepsilon = 0.25$ are substantially different, due to qualitatively different dynamics, but the variation of the entropy is small in both cases, thus showing no chaotic itinerancy.

\section{Local permutation entropy}
\label{sec:perm}

The analysis of the variance of local Shannon entropy in a time series, described in Section~\ref{sec:entropy}, allows one to find parameters for which systems may potentially exhibit the phenomenon of chaotic itinerancy.
However, this method is not universally effective. One can think of examples in which high local Shannon entropy is obtained even though the sequence actually exhibits ordered behavior; this can happen, for example, if the values are locally evenly distributed. Because of that, we additionally propose to use permutation entropy, which allows one to distinguish certain cases of ordered behavior that are not captured by Shannon entropy.

Permutation entropy is a measure of time series complexity based on ordinal patterns of successive values \cite{Bandt2002-xy}. Instead of considering the exact values of the data points, ordinal patterns capture the relative ordering within short subsequences.

Two key parameters in the computation of the permutation entropy are the pattern length $d$ and the time delay $\tau$. The length of the ordinal patterns defines how many consecutive or delayed values are grouped into each vector vector. The time delay $\tau$ sets the time interval between successive elements in each vector.

\emph{Permutation entropy} of a time series $X = (x_i)_{i=1}^n$ is defined as the Shannon entropy of the distribution of ordinal patterns of length $d$:
\begin{equation}
\mbox{PE}(X) = - \sum_{i=1}^{d!} p(\pi_i) \log_2 p(\pi_i),
\end{equation}
where $p(\pi_i)$ is the observed probability of appearance of the ordinal pattern $\pi_i$ in the time series $X$.

Analogously to the local Shannon entropy, we define \emph{local permutation entropy} for a given point in the sequence $X = (x_i)_{i=1}^n$ as follows:
\begin{equation}
\mbox{PE}_{\text{local}}(j) = \mbox{PE}(X_{j-L,j+L}),
\end{equation}
where $X_{j-L,j+L}$ denotes the fragment of the sequence $X$ that includes $2L + 1$ consecutive elements from $x_{j-L}$ to $x_{j+L}$, inclusive.

The optimal choice of $\tau$ depends on a particular system considered. For a discrete-time dynamical system, the natural choice for the time delay is $\tau=1$. However, if one considers a time-discretization of a system with continuous time (a flow), then a short time step may not be sufficient for the discrete trajectories to reflect dynamically relevant changes ($x_i$ may be very close to $x_{i+1}$), and then larger values of $\tau$ may be desired, corresponding to time after which $x_i$ an $x_{i+\tau}$ become separated in the phase space. For example, one may choose $\tau$ corresponding to the first local minimum of mutual information between $x_i$ and $x_{i+\tau}$, as suggested in \cite{Fraser1986,Myers2019}.

The choice of suitable values of $L$ and $d$ must be coordinated together. The number of different ordinal patterns of length $d$ is $d!$, so $L$ should be large enough to allow one to gather enough statistics on the appearance of all the $d!$ patterns in segments of length $2L+1$; for example, $L > 2 d!$ might be a reasonable request. Choosing a larger value of $d$ provides a finer insight into the dynamics, offering a higher number of possible sequences, but decreasing the potential number of their appearances. With $L=100$, choosing $d=3$ makes the average number of appearances of each pattern approximately $67$, while choosing $d=4$ decreases this number to almost $17$, which is still reasonable.

\begin{figure}[htbp]
\centering
\includegraphics[width=0.85\textwidth]{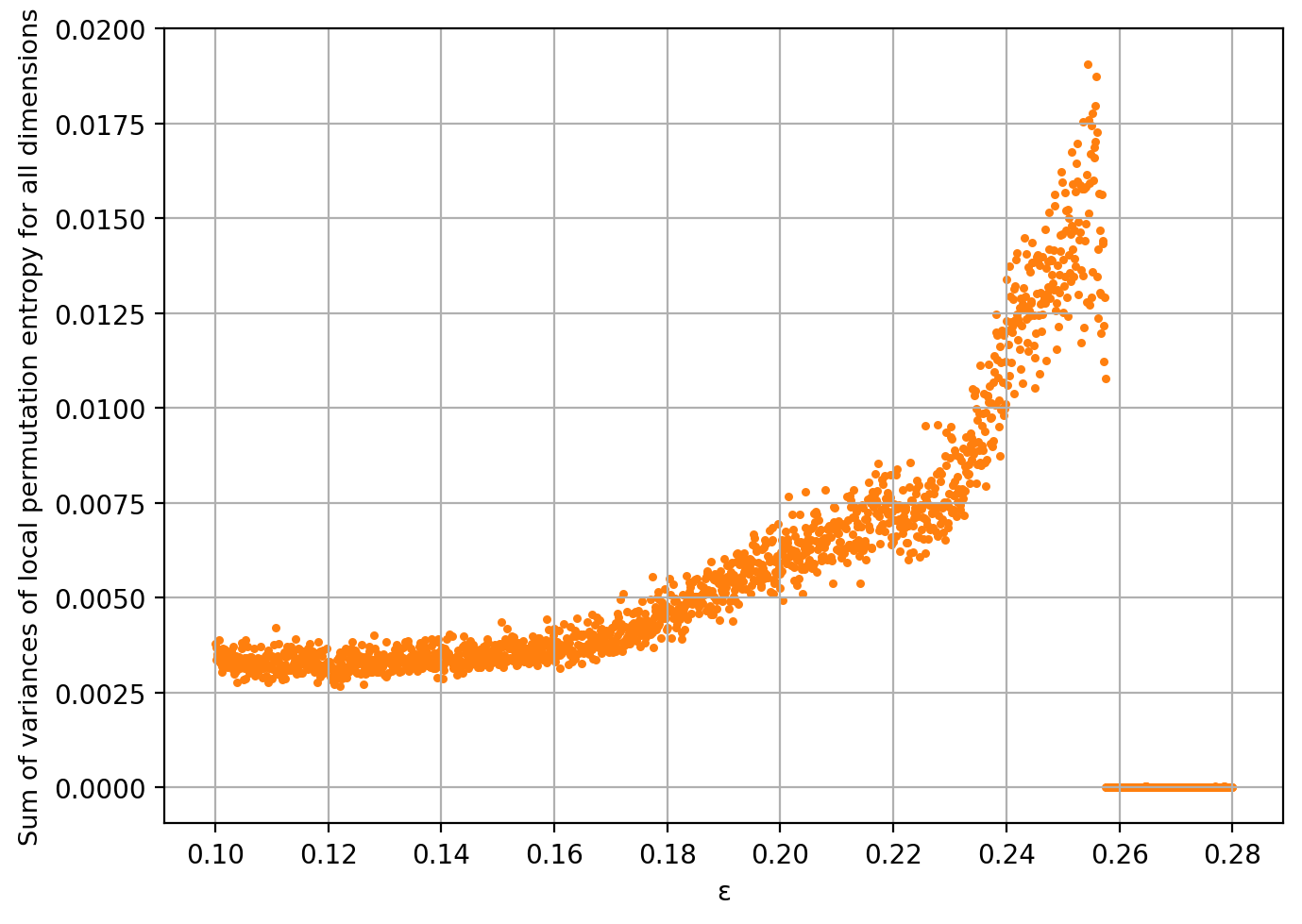}
\caption{Sum of variances of local permutation entropy for all the coordinates of points on a segment of a sample trajectory as a function of $\varepsilon$, computed for the GCM model with $N=3$ and $a=2$.}
\label{sum_perm_ent}
\end{figure}

To obtain better insight into the role of the choice of $d$ and $L$, we tested $d \in \{2, 3, 4, 5, 6\}$ for all $L \in \{50, 100, 200, 500\}$ and obtained similar results in all the cases except for $d = 2$, where we were not able to see the peek shown in Figure~\ref{sum_perm_ent} (discussed below). It follows that ordinal patterns of length $2$ do not have enough discriminative power to provide sufficient information about the dynamics. On the other hand, all the other values of $d$ yielded similar information in our case, so we choose to work with $d=3$.

Let us consider ordinal patterns of length $d=3$ with time delay $\tau=1$, which means that the time series is examined in overlapping segments of three consecutive values. Let us fix $L=100$ as previously. For each segment of length $d=3$, we determine the relative order of the three points. The measured frequency of appearance of each possible ordering is then used to compute the Shannon entropy of the distribution of these patterns.

Computed values of the variance of local permutation entropy in the GCM model as a function of $\varepsilon$ with $N=3$ and $a=2$ are shown in Figure \ref{sum_perm_ent}. The values of $\varepsilon$ ranging from $0.1$ to $0.28$ are considered, with the step of $0.0001$. The results are consistent with the results for local Shannon entropy shown in Figure~\ref{sum_ent}. One can see gradual increase in the local permutation entropy with the increase in $\varepsilon$ until $\varepsilon \approx 0.25$, which corresponds to gradual transition from classic chaos to chaotic itinerancy, and then a sudden drop to $0$ corresponding to a bifurcation into an ordered state, as discussed in Section~\ref{sec:entropy}. These three types of dynamics are shown in Figure~\ref{different_eps}.

\begin{figure}[htbp]
\centering
\includegraphics[width=1\textwidth]{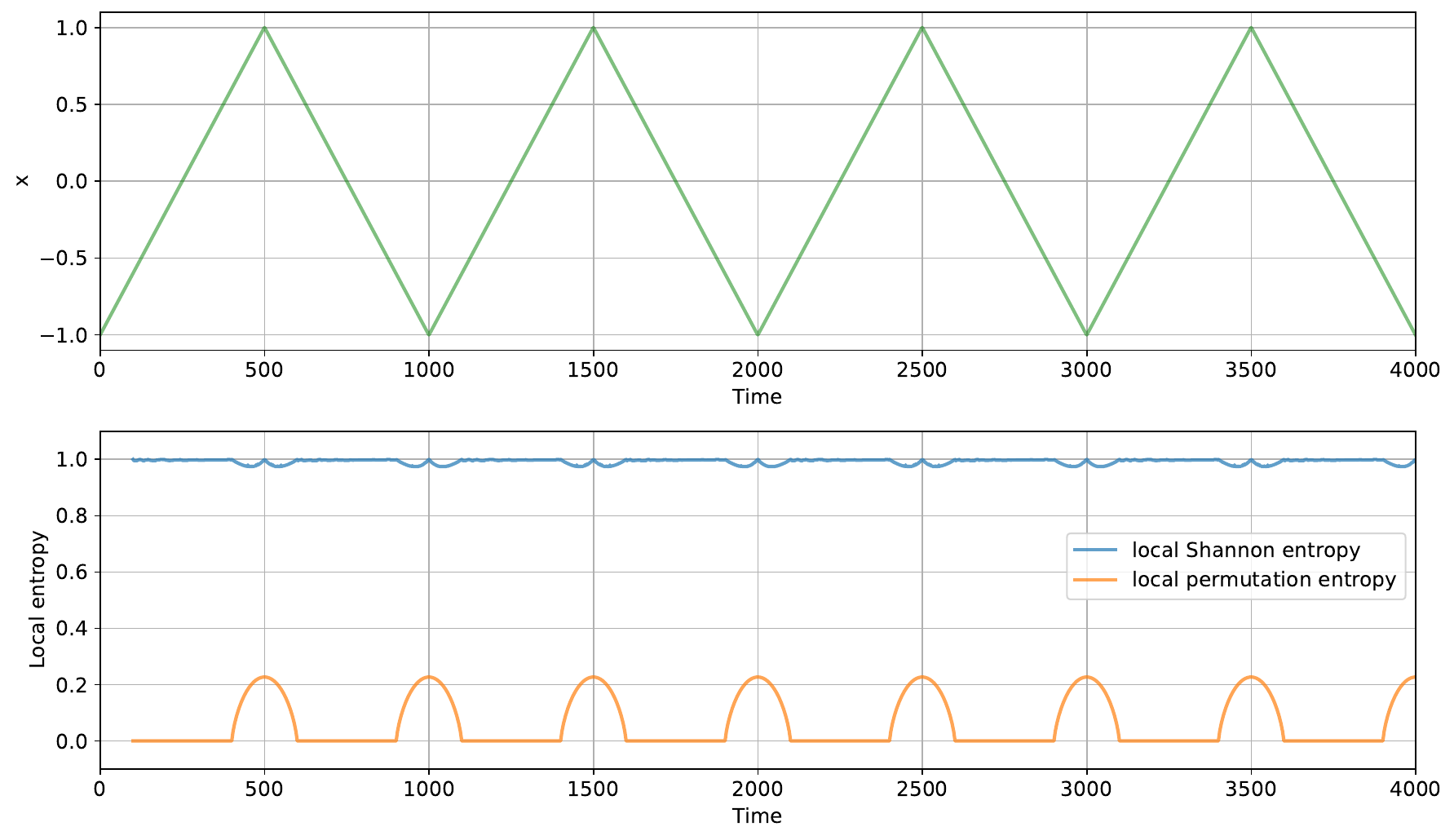}
\caption{Example of an ordered time series for which local Shannon entropy is close to its maximum, while local permutation entropy attains lower values. The entropy values are divided by the maximum possible value in each case.}
\label{example_permutation}
\end{figure}

Let us now illustrate the advantage of using local permutation entropy over local Shannon entropy by constructing an artificial example of a very specific time series that we show in Figure \ref{example_permutation}. The upper part of the figure shows the time series that steadily increases and then decreases; this behavior is then repeated periodically. The bottom part of the figure shows local Shannon entropy computed at each point (the blue line), which is close to its maximum, while permutation entropy (the orange line) detects the underlying order and attains very low values.

However, it should be noted that there may be other situations in which neither Shannon entropy nor permutation entropy fulfills the intended role.
Nevertheless, local permutation entropy may be considered a valuable addition to local Shannon entropy for the purpose of detecting possible existence of chaotic itinerancy.

\section{Clustering}
\label{sec:dbscan}

Attractor ruins resemble traditionally understood attractors because they attract trajectories, even though a trajectory typically stays in their vicinity only for a limited period of time and eventually leaves. When analyzing a single trajectory, we expect that the points of such a trajectory form dense clusters around attractor ruins visited by the trajectory in the phase space. Therefore, we propose to use a density-based clustering algorithm to identify the attractor ruins.

HDBSCAN is a clustering algorithm that extends the classic density-based clustering algorithm DBSCAN~\cite{Ester1996-dbscan} by building a hierarchy of clusters based on density~\cite{Campello2013}. It only requires one main parameter: minimum cluster size---the smallest number of points that a cluster should contain.
The algorithm works by computing core distances for each point, building a mutual reachability graph, and then constructing a minimum spanning tree. It creates a hierarchy of clusters by progressively removing edges based on density and selects the most stable clusters from this hierarchy. HDBSCAN is capable of detecting clusters with varying densities and classifies scattered points that do not belong to any cluster as noise.

Let us now describe an application of the HDBSCAN algorithm to the detection of attractor ruins in one of the cases suspected of exhibiting chaotic itinerancy, as suggested by the increased value of local Shannon entropy variance for $\varepsilon=0.2574$ in the plot shown in Figure \ref{sum_ent}. The investigated model is GCM with $N = 3$ and $a = 2$.

We take a pseudo-random initial condition and compute consecutive iterations to generate a trajectory $(x_n)$ consisting of $K=40000$ points. We discard the first $K_0=20000$ iterations and analyze the segment consisting of $K-K_0 = 20000$ points, starting after the $K_0$ initial iterations that we consider necessary to allow the dynamics to settle down on a global attractor.

\begin{figure}[htbp]
\centering
\includegraphics[width=0.9\textwidth]{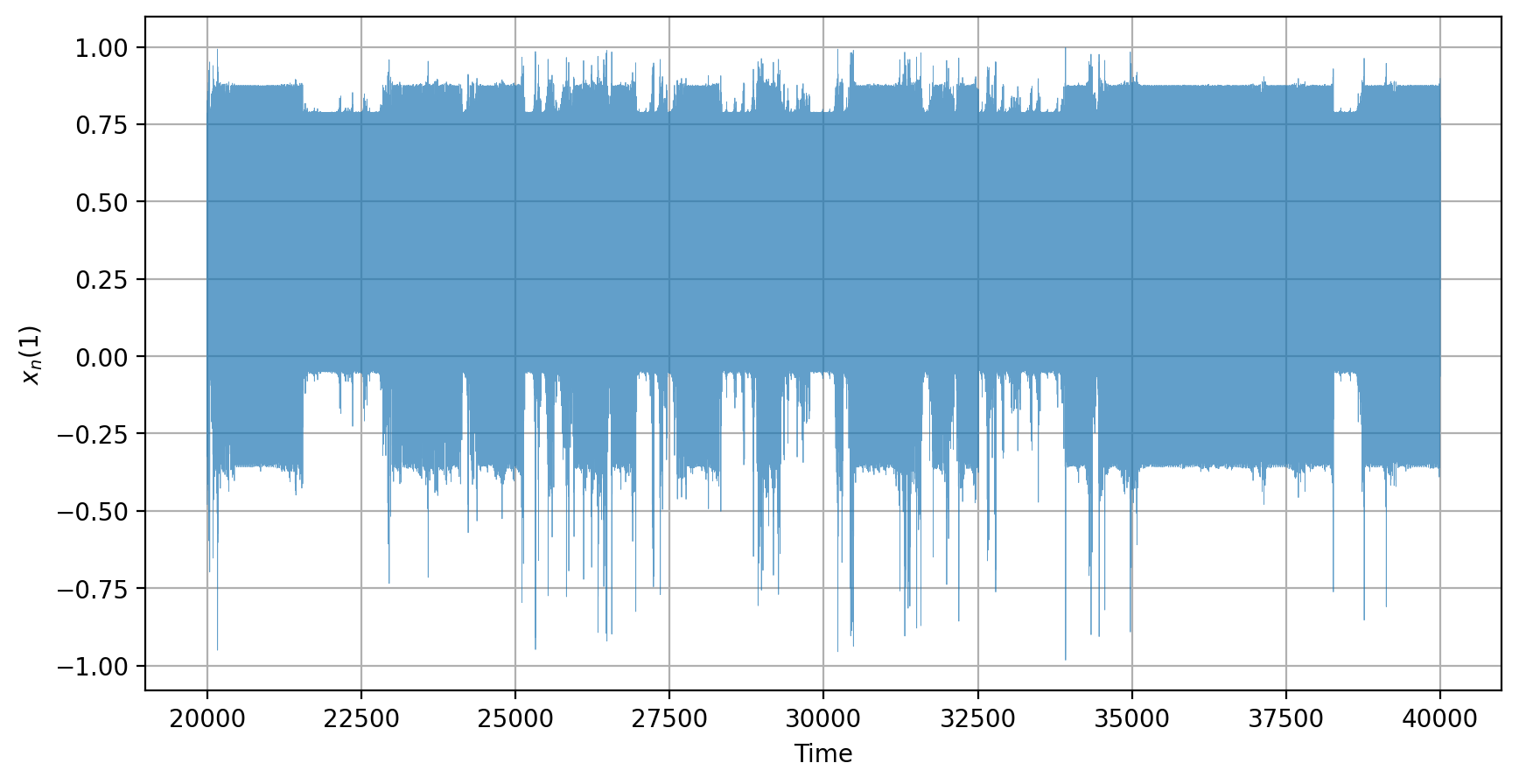}
\caption{Time series $x_n(1)$ for a sample trajectory in the GCM model with $N=3$, $a=2$ and $\varepsilon = 0.2574$.}
\label{time_series_gcm}
\end{figure}

The time series for $x_n(1)$, illustrated in Figure \ref{time_series_gcm} for $n \in [K_0,K]$, reflects the expected dynamics, characterized by a sequence of ordered and chaotic phases. One can see intervals of various lengths with nearly constant amplitude within the interval, most prominently the wide interval between 35000 and 37500. Such intervals correspond to ordered dynamics, with the trajectory oscillating in the vicinity of an attractor ruin. The different amplitudes observed for such intervals correspond to different attractor ruins. One can also see intervals characterized by high irregularity of the amplitude. Such intervals correspond to chaotic transitions between the attractor ruins. We remark that it is sufficient to analyze the time series of a single variable in this case because the remaining variables exhibit similar dynamics.

HDBSCAN has a single parameter that controls its action: the minimum requested cluster size. In order to choose an appropriate value of this parameter, we propose to try a few different numbers (e.g. between $50$ and $1000$) and choose the best one based on the silhouette score, a metric often used to evaluate how well clusters are formed \cite{Rousseeuw1987-silhouettes}.
In our case, we computed the silhouette score for clusters found using HDBSCAN with the minimum cluster sizes of $50$, $100$, $150$, $200$, $300$, $400$, $500$, $600$. We obtained the best score in the case of $M=300$, so this is the parameter value that we chose for further considerations.

\begin{figure}[htbp]
\centering
\includegraphics[width=0.7\textwidth]{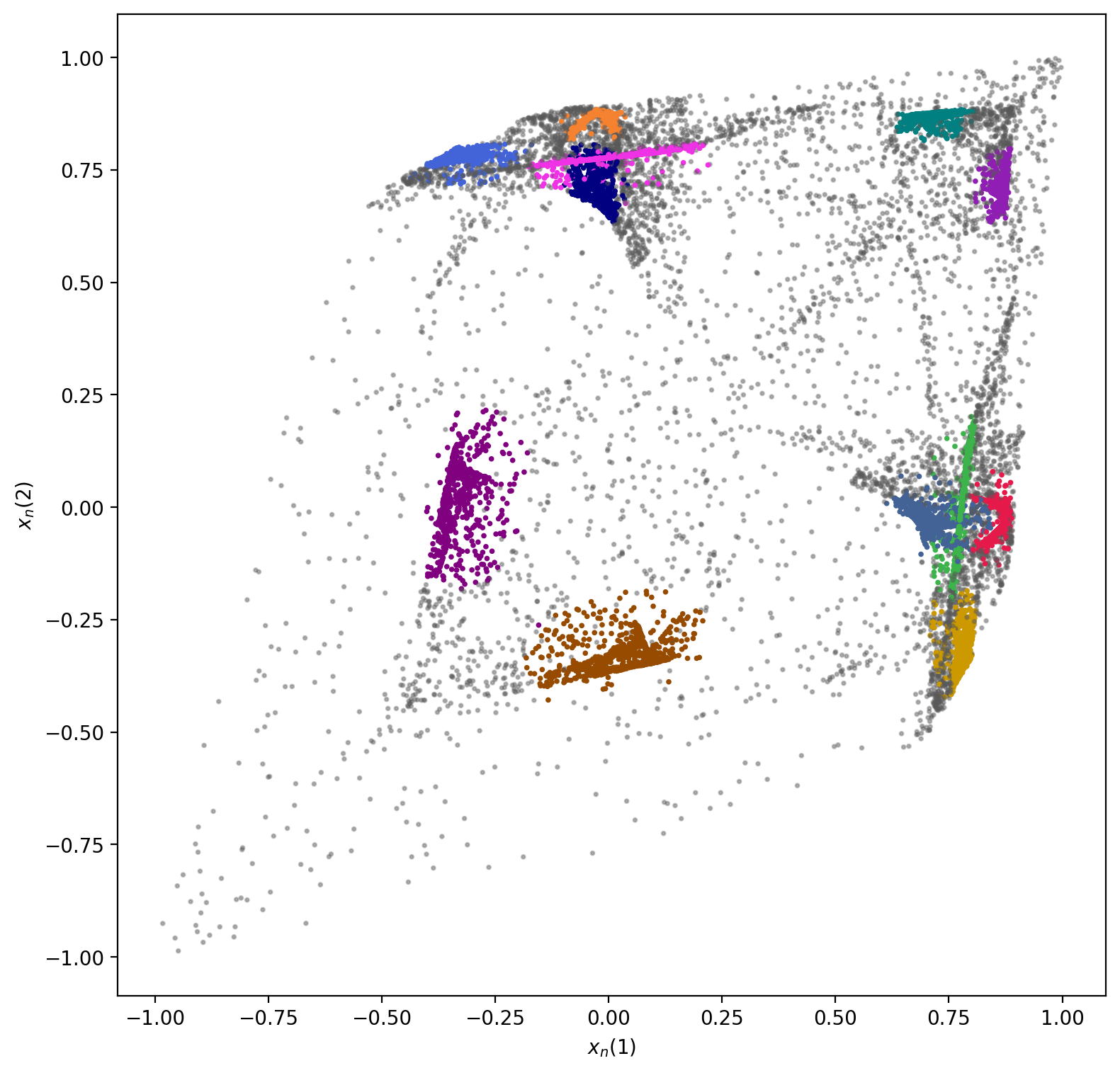}
\caption{Clusters obtained using HDBSCAN in a sample trajectory in the GCM model with $N=3$, $a=2$ and $\varepsilon = 0.2574$, projected onto the $(x(1),x(2))$ plane.}
\label{clusters_gcm}
\end{figure}

We apply the HDBSCAN algorithm with the minimum cluster size set to $M=300$. Figure \ref{clusters_gcm} depicts the clusters obtained in this way. The algorithm successfully identified $12$ clusters and left the irregularly distributed scattered points in the space classified as noise.

The number of points within each cluster is provided in Table \ref{tab:gcmclusters}, together with the number of points that were not assigned to any of the clusters (noise). Note that noise constitutes approximately $29\%$ of all the points. There are $12$ clusters containing around $1000$ points each.

\begin{table}[htbp]
\centering
\setlength{\tabcolsep}{4pt} 
\footnotesize
\begin{tabular}{c|ccccccccccccc}
\textbf{Cluster} & \textbf{noise} & \textbf{0} & \textbf{1} & \textbf{2} & \textbf{3} & \textbf{4} & \textbf{5} & \textbf{6} & \textbf{7} & \textbf{8} & \textbf{9} & \textbf{10} & \textbf{11} \\
\hline
\textbf{N.\ points} & 5731 & 844 & 836 & 1557 & 1562 & 1251 & 1227 & 1550 & 1571 & 780 & 783 & 1153 & 1155 \\
\end{tabular}
\vspace{6pt}
\caption{Number of points in each cluster shown in Figure~\ref{clusters_gcm}.}
\label{tab:gcmclusters}
\end{table}

\begin{table}[htbp]
\centering
\footnotesize
\setlength{\tabcolsep}{4pt}
\begin{tabular}{c|rrrrrrrrrrrrr}
 & \textbf{noise} & \textbf{0} & \textbf{1} & \textbf{2} & \textbf{3} & \textbf{4} & \textbf{5} & \textbf{6} & \textbf{7} & \textbf{8} & \textbf{9} & \textbf{10} & \textbf{11} \\
\hline
\textbf{noise}  & 5151 & 71  & 63  & 68  & 73  & 113  & 89  & 34  & 43  & 6   & 5   & 8   & 6 \\
\textbf{0}      & 70   & 0   & 0   & 0   & 0   & 0    & 0   & 0   & 0   & 774 & 0   & 0   & 0 \\
\textbf{1}      & 58   & 0   & 0   & 0   & 0   & 0    & 0   & 0   & 0   & 0   & 778 & 0   & 0 \\
\textbf{2}      & 40   & 0   & 0   & 0   & 0   & 0    & 0   & 1516& 1   & 0   & 0   & 0   & 0 \\
\textbf{3}      & 35   & 0   & 0   & 0   & 0   & 0    & 0   & 0   & 1527& 0   & 0   & 0   & 0 \\
\textbf{4}      & 101  & 0   & 0   & 0   & 0   & 0    & 0   & 0   & 0   & 0   & 0   & 1   & 1149 \\
\textbf{5}      & 83   & 0   & 0   & 0   & 0   & 0    & 0   & 0   & 0   & 0   & 0   & 1144& 0 \\
\textbf{6}      & 61   & 0   & 0   & 0   & 1489& 0    & 0   & 0   & 0   & 0   & 0   & 0   & 0 \\
\textbf{7}      & 82   & 0   & 0   & 1489& 0   & 0    & 0   & 0   & 0   & 0   & 0   & 0   & 0 \\
\textbf{8}      & 7    & 0   & 773 & 0   & 0   & 0    & 0   & 0   & 0   & 0   & 0   & 0   & 0 \\
\textbf{9}      & 10   & 773 & 0   & 0   & 0   & 0    & 0   & 0   & 0   & 0   & 0   & 0   & 0 \\
\textbf{10}     & 15   & 0   & 0   & 0   & 0   & 1138 & 0   & 0   & 0   & 0   & 0   & 0   & 0 \\
\textbf{11}     & 17   & 0   & 0   & 0   & 0   & 0    & 1138& 0   & 0   & 0   & 0   & 0   & 0 \\
\end{tabular}
\vspace{6pt}
\caption{Transition matrix of the analyzed trajectory between the clusters found. The number of points transitioning from cluster $i$ to cluster $j$ in one step is shown in the $i$-th row and $j$-th column.}
\label{tab:gcmtransition}
\end{table}

\begin{figure}[htbp]
\centering
\includegraphics[width=0.7\textwidth]{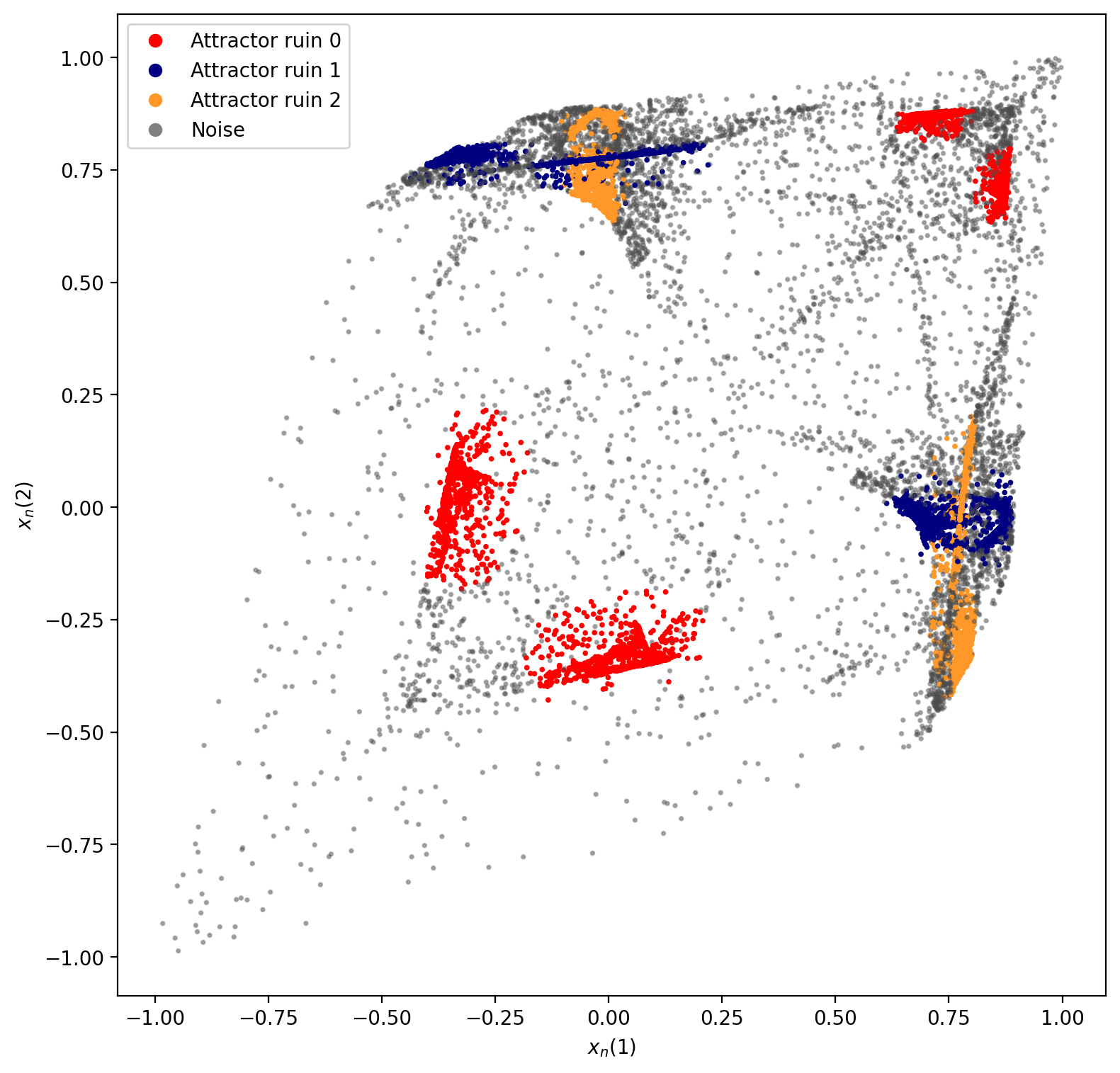}
\caption{Attractor ruins obtained from clusters shown in Figure~\ref{clusters_gcm} by merging quadruplets of clusters corresponding to $4$-periodic sequences determined by the analysis of transitions shown in Table~\ref{tab:gcmtransition}.}
\label{attractor_ruins}
\end{figure}

The trajectory exhibits characteristic movement between specific clusters. It can be summarized by means of a transition matrix $T$ whose rows and columns correspond to the clusters, with the first row and the first column additionally corresponding to the noise, defined as follows:
\[
T_{ij} = \mbox{card} \{n \in [K_0,K-1] : x_n \in C_i, x_{n+1} \in C_j\},
\]
where $C_i$ denotes the cluster indexed by the integer number $i$ (starting with $0$), or the set of scattered points (noise). In the transition matrix $T$ shown in Table~\ref{tab:gcmtransition} almost all the points in $C_0$ transition in the next step to $C_8$, then we can see a large number of points moving from $C_8$ to $C_1$, from $C_1$ to $C_9$ and from $C_9$ to $C_0$. This corresponds to a periodic orbit with period~$4$. Therefore, we merge these four clusters into one set that is an approximation of an attractor ruin; we shall call it an \emph{attractor ruin} for short. A similar observation applies to clusters $C_2$--$C_6$--$C_3$--$C_7$ and $C_4$--$C_{11}$--$C_5$--$C_{10}$. The merging of the three quadruplets of the clusters results in three attractor ruins, indicated in Figure~\ref{attractor_ruins}. As a result, the attractor ruins consist of clusters between which the trajectory transitions in an ordered manner. To emphasize the contrast between the ordered dynamics in the attractor ruins and the unordered dynamics outside of them, we shall use the term ``chaotic transition state'' to indicate the set of points classified as noise.

\section{Chaotic itinerancy}
\label{sec:chaotic}

When a trajectory enters one of the attractor ruins found in the system, it stays in its vicinity for some time before ultimately departing from this state. The number of such transitions from each attractor ruin to the chaotic transition state is shown in the first column of Table \ref{tab:ruins_transition}, where one should keep in mind that the three attractor ruins consist of four clusters each, so in fact the contents of Table~\ref{tab:ruins_transition} can be derived directly from Table~\ref{tab:gcmtransition}.

\begin{table}[htbp]
\centering
\footnotesize
\setlength{\tabcolsep}{6pt}
\begin{tabular}{c|rrrr}
 & \textbf{noise} & \textbf{0} & \textbf{1} & \textbf{2} \\
\hline
\textbf{noise} & 5151 & 145 & 218 & 216 \\
\textbf{0} & 145 & 3098 & 0 & 0 \\
\textbf{1} & 218 & 0 & 6022 & 0 \\
\textbf{2} & 216 & 0 & 0 & 4570 \\
\end{tabular}
\vspace{6pt}

\caption{Transition matrix of the analyzed trajectory between the attractor ruins found (labeled 0, 1 and 2) and the chaotic transition state (noise). The number of points transitioning from attractor ruin $i$ (or noise) to attractor ruin $j$ (or noise) in one step is shown in the row labeled $i$ and the column labeled $j$, where $i,j \in \{\mbox{noise}, 0, 1, 2\}$.}
\label{tab:ruins_transition}

\end{table}

\begin{figure}[htbp]
\centering
\includegraphics[width=1\textwidth]{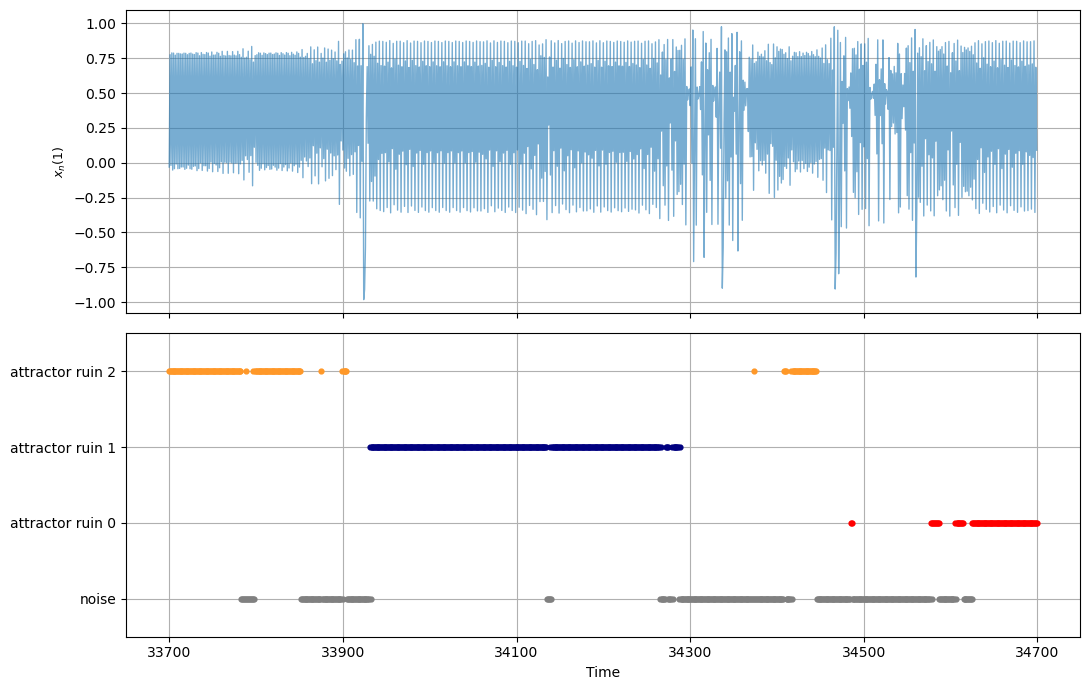}
\caption{Time series of the first coordinate (top graph) and attractor ruin membership (bottom graph) of a segment of a sample trajectory in the GCM with $N=3$, $a=2$ and $\varepsilon=0.2574$, wandering between the three attractor ruins found in the system.}
\label{time_series_clusters}
\end{figure}

A segment of the analyzed time series is shown in Figure \ref{time_series_clusters} along with attractor ruin membership for each point. It can be observed that irregular (chaotic) behavior of the time series corresponds to points classified as noise, whereas ordered motion can be associated with the assignment to one of the attractor ruins.

An interesting observation is that the variation of $x_n(1)$ shown in Figure~\ref{time_series_clusters} in the segments of the time series assigned to attractor ruin 2 is different than the variation of $x_n(1)$ in attractor ruins 0 and 1. Indeed, one can confirm in Figure~\ref{attractor_ruins} that the projection of attractor ruins 0 and 1 onto the first (horizontal) coordinate $x_n(1)$ would extend over a wider range than the projection of attractor ruin 2. This explains the observed difference in the amplitude of the time series $x_n(1)$ between these attractor ruins.

Another observation that we would like to draw attention to is that the range of coordinates of the points classified as noise shown in Figure~\ref{attractor_ruins} is wider than the range of coordinates in the attractor ruins, especially regarding negative values. This is clearly reflected in Figure~\ref{time_series_clusters} where one can see occasional ``spikes'' pointing downwards in the time series plot in the segments classified as noise.

After assigning each point on a trajectory to a specific attractor ruin (or to noise), additional information on the dynamics can be obtained by determining the number of visits in each attractor ruin and calculating the time spent in the attractor ruin during each visit. Results obtained in our case are shown in Table \ref{average_time}.

\begin{table}[htbp]
\centering
\begin{tabular}{c|cccc}
\textbf{Attractor ruin} & \textbf{noise} & \textbf{0} & \textbf{1} & \textbf{2} \\
\hline
\textbf{Average time} & 9.88 & 22.37 & 28.62 & 22.16 \\
\textbf{Median time} & 4 & 1 & 1 & 1  \\
\textbf{Std. dev. of time} & 15.10 & 72.99 & 147.79 & 64.46 \\
\textbf{Number of visits} & 580 & 145 & 218 & 216 \\
\end{tabular}
\vspace{6pt}
\caption{Average and median time spent in each attractor ruin or in the chaotic transition state (noise) by the analyzed trajectory in the GCM model. Additionally, standard deviation of the visit times is provided, as well as the number of visits encountered.}
\label{average_time}
\end{table}

We remark that the quantities shown in Table~\ref{average_time} such as the time spent in an attractor ruin $A$ by the trajectory $(x_n)$ can be formally defined as follows. We say that the trajectory enters $A$ at the time point $k$ if $x_{k-1} \notin A$ and $x_{k} \in A$. Denote the set of such time points between $K_0$ and $K$ by $E(A,K_0,K)$. Then the time spent in $A$ associated with a time point $k \in E(A,K_0,K)$ is defined as follows:
\[
t_k = \mbox{max} \{l>0 : x_{i} \in A \text{ for all } i = k, \ldots, k+l-1\}.
\]
The average, median, or standard deviation of the time spent in the attractor ruin $A$ is then defined, respectively, as the average, median, or standard deviation of the sequence $(t_k)_{k \in E(A,K_0,K)}$. The number of visits to $A$ is defined as $\mbox{card}\, E(A,K_0,K)$.
The corresponding notions regarding the chaotic transition state $C_{\mbox{\scriptsize noise}}$ can be defined by replacing $A$ with $C_{\mbox{\scriptsize noise}}$ in the definitions above. In particular, the first entry of Table~\ref{average_time} shows the average time spent in the chaotic transition state.

It should be noted that clustering may not be perfectly aligned with the actual attractor ruins, as there might be certain points classified in a suboptimal way. Indeed, in the attractor ruin assignment plot (Figure \ref{time_series_clusters}), isolated points appear that are assigned in a different way than surrounding points in the series. Additional verification shows that they are located in close proximity to the identified attractor ruins. This is likely due to the fact that HDBSCAN favors persistent dense regions over transitional density areas. These points can significantly affect the estimated time spent in an attractor ruin. Indeed, the median of these times is mostly $1$, which illustrates the massive scale of this phenomenon and prompts the need to fix the problem. We propose to assign such points to attractor ruins or to the chaotic transition state, according to the surrounding points in the time series.

Table \ref{average_time_merged}
shows the numbers of visits and their times after adjusting the assignments of the isolated points as discussed above. With this correction, the average visiting times are considerably higher, and we feel that they reflect the actually observed dynamics in a more accurate way. The numbers of visits are lower, and the medians now have meaningful values.

\begin{table}[htbp]
\centering
\begin{tabular}{c|cccc}
\textbf{Attractor ruin} & \textbf{noise} & \textbf{0} & \textbf{1} & \textbf{2} \\
\hline
\textbf{Average time} & 26.71 & 59.91 & 74.30 & 53.74\\
\textbf{Median time} & 17 & 11 & 8 & 8  \\
\textbf{Std. dev. of time} & 29.06 & 115.96 & 232.69 & 93.63 \\
\textbf{Number of visits} & 224 & 53 & 83 & 87 \\
\end{tabular}
\vspace{6pt}
\caption{Average and median time spent in each attractor ruin or in the chaotic transition state (noise) by the analyzed trajectory in the GCM model after applying the correction described in the text. Additionally, standard deviation of the visit times is provided, as well as the number of visits encountered.}
\label{average_time_merged}
\end{table}

In order to assess whether the itinerancy is chaotic, we analyze the randomness of the sequence of consecutive attractor ruins visited by the trajectory, as well as the sequence of times spent in the attractor ruins. The sequence of consecutive attractor ruins visited consists of elements where the first element is the label of the initial attractor ruin visited by the trajectory, and the subsequent elements are the labels of the attractor ruins the trajectory enters after leaving the previous attractor ruin. The elements of the sequence of times spent in the attractor ruins represent the number of time steps from entry to exit for each successive attractor ruin visited by the trajectory. We propose three tests to assess the randomness of the observed wandering:

\begin{enumerate}
    \item \label{test:LB} Ljung--Box test that determines whether a time series exhibits significant autocorrelation \cite{Ljung1978}. We propose to perform this test with $10$ lags.
    \item \label{test:ADF} Augmented Dickey--Fuller test that checks for the presence of a unit root in a time series, indicating whether the series is non-stationary \cite{Dickey1979}.
    \item \label{test:OBD} Wald--Wolfowitz runs test \cite{Wald1940} that can be used to assess randomness in a binary time series, or O'Brien--Dyck runs test that is more suitable for categorical data with more than two categories \cite{OBrien1985}.
\end{enumerate}

Tests \eqref{test:LB} and \eqref{test:ADF} are used to assess the randomness of the sequence of times spent in the attractor ruins, while test \eqref{test:OBD} is applied to the sequence of consecutive attractor ruins visited.

Table \ref{tests_gcm} contains the results of tests \eqref{test:LB}, \eqref{test:ADF} and \eqref{test:OBD} for the analyzed trajectory in the GCM model. Note that since in this case we have three attractor ruins, we use O'Brien--Dyck runs test instead of Wald--Wolfowitz. We apply all the tests for the corrected point labels obtained after assigning isolated points to attractor ruins or to the chaotic transition state, according to the surrounding points in the time series, as explained earlier. The statistical tests in this paper are carried out using the significance level $\alpha = 0.05$. We use the implementation of test \eqref{test:OBD} provided on~\cite{code_runs_test}, see~\cite{Walk2014}.

\begin{table}[htbp]
\centering
\begin{tabular}{l|c|l}
\small
\textbf{Test} & \textbf{p-value} & \textbf{Interpretation} \\
\hline
Ljung-Box Test & 0.784 & No detectable autocorrelation \\
Augmented Dickey-Fuller Test & 0.000 & No signs of non-stationarity \\
O'Brien-Dyck Runs Test & 0.659 & No evidence of non-randomness \\
\end{tabular}
\vspace{6pt}
\caption{Summary of statistical tests applied to assess randomness of chaotic itinerancy observed in the GCM model, including corresponding p-values and interpretations.}
\label{tests_gcm}
\end{table}

The p-values shown in Table~\ref{tests_gcm} indicate that the sequence of times spent in the attractor ruins shows no evidence of autocorrelation or non-stationarity, and that the sequence of consecutive attractor ruins visited by the trajectory lacks any regular pattern. Therefore, we conclude that the transitions between attractor ruins are random, confirming that in this case we observe chaotic itinerancy.

In addition to the analysis of the dynamical features related directly to the attractor ruins, one can also ask questions about the dynamics in the chaotic transition state itself. Indeed, one can see in Figure \ref{attractor_ruins} that the distribution of points in the chaotic transition state is not uniform. There are some regions with higher density than others. However, they do not contain a sufficient number of points to be identified by the HDBSCAN algorithm as separate clusters, and the density is not high enough to make them part of the near-by clusters. Many of these regions are located near the attractor ruins and form some sort of whiskers or protrusions. We checked that some of the points in these regions exhibit dynamics resembling ordered behavior in the attractor ruins, moving from the vicinity of one part of an attractor ruin to the vicinity of another part of the same attractor ruin, for some time following the periodic dynamics among the clusters that form the attractor ruins. Thus, the dynamics in the chaotic transition state are not entirely chaotic, yet it clearly differs from transitions occurring within attractor ruins. We additionally performed tests \eqref{test:LB} and \eqref{test:ADF} on the sequence of times spent in the chaotic transition state. The results indicate absence of autocorrelation and non-stationarity of this sequence. These results allow us to conclude that the dynamics in the chaotic transition state is chaotic.

\section{Comprehensive analysis of chaotic itinerancy in the GCM model}
\label{sec:hdbscan}

Based on the methods used for the analysis of clusters and attractor ruins described in Sections \ref{sec:dbscan} and~\ref{sec:chaotic}, we propose an automated method for scanning a wide range of parameters in search of chaotic itinerancy in a dynamical system. We show its application to the GCM model, but we emphasize the fact that the method is general and can be applied to a variety of systems, possibly after adjusting the various hyperparameters of the method like $K$, $K_0$ or $M$.

The following procedure is repeated for all combinations of parameters of interest. First, $K=40000$ iterations of a sample trajectory are generated and the first $K_0=20000$ iterates of the trajectory are discarded to allow the dynamics stabilize. Next, HDBSCAN is applied with the minimum cluster size set to $M=300$. Cluster pairs with more than $80\%$ of all transitions that point from one cluster to the other are identified. Clusters that exhibit these dominant transitions, including cyclical transitions that involve several clusters, are then merged together into larger sets of points that we call attractor ruins. Similarly to what we described in Section~\ref{sec:chaotic}, the assignments of isolated points are adjusted next, which results in slightly larger attractor ruins. We say that chaotic itinerancy is not found if fewer than two attractor ruins remain or the proportion of points classified as noise is below $10\%$ or above $90\%$. Otherwise, we say that the system passes the first stage of verification of chaotic itinerancy.

\begin{figure}[htbp]
\centering
\includegraphics[width=1\textwidth]{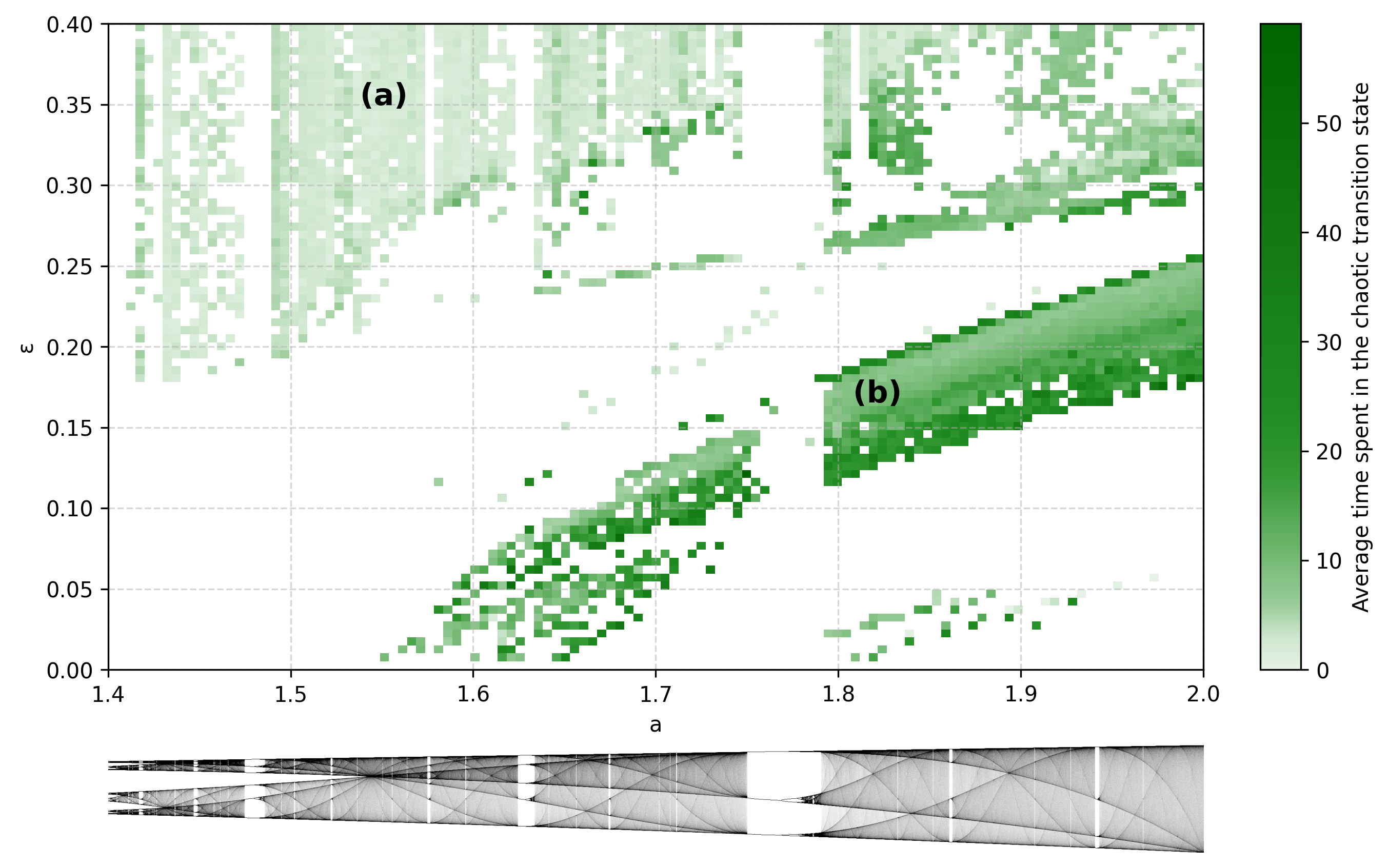}
\caption{Average time spent in the chaotic transition state in the GCM system with $N=3$ and a selection of different values of $a \in [1.4,2]$ and $\varepsilon \in [0.005,0.4]$. The bifurcation diagram of the logistic map is shown for reference. The regions (a) and (b) are discussed in the text.}
\label{grid}
\end{figure}

Figure \ref{grid} shows a grid of $121 \times 80$ parameters $a \in [1.4, 2]$ and $\varepsilon \in [0.005, 0.4]$. Different shades of green indicate the parameters for which a sample trajectory generated for the GCM system with these parameter values and $N=3$ passed the first verification stage of the chaotic itinerancy described above. The parameter values for which this verification failed are shown in white. The intensity of the color indicates the average time spent in the chaotic transition state according to the scale shown to the right of the plot. This is the quantity shown in the very first entry of Table~\ref{average_time_merged}. The bifurcation diagram of the logistic map is shown below the plot for reference.

The first characteristic feature that can be noticed in Figure~\ref{grid} is the vertical white stripes that correspond to periodic windows of the logistic map. The widest of such stripes is clearly visible between $a = 1.75$ and $a = 1.8$. The inability to find suitable attractor ruins for these parameter values confirms that the presence of chaotic itinerancy should not be expected if the one-dimensional maps that are coupled together exhibit attracting periodic orbits.

One can also notice two major regions in the figure in which at least two attractor ruins were successfully identified and the level of noise was acceptable to pass the first stage of verification of chaotic itinerancy. The first region, labeled (a), spans in the top part of the figure, for $\varepsilon$ above approximately $0.25$. The second region, labeled (b), forms a distinct slanted band that spans from $(a,\varepsilon) \approx (1.6, 0)$ at the bottom of the figure to $(a,\varepsilon) \approx (2, 0.2)$ on the right of the figure. There is a clearly visible belt between these regions in which no relevant attractor ruins were found for the vast majority of parameters there. We would like to draw attention to the predominance of pale shades of green in (a), as opposed to the dark green that often appears in (b). Recall that the intensity of the color indicates the average time spent in chaotic transition state. Since the average times in region (b) are mostly above $10$, this suggests that there is enough room for chaotic dynamics during the transitions between attractor ruins, and therefore it is reasonable to expect the existence of chaotic itinerancy for these parameters. However, for parameters in region (a), the average number of iterations in the chaotic transition state rarely exceeds $5$, which indicates very short jumps between visiting the vicinity of attractor ruins.

The shape of the two regions (a) and (b) shown in Figure~\ref{grid} resembles the shape of the rough phase diagram sketched by Kaneko in 1990 on the basis of visual inspection of the results of his numerical experiments \cite[Figure~3]{Kaneko1990-fj}, in which the phases described in Section~\ref{sec:gcm} were distinguished. The location of region (b) that we observe in Figure~\ref{grid} corresponds to what Kaneko called the intermittent phase located at the boundary between the turbulent and ordered phases.

\begin{figure}[htbp]
\centering
\includegraphics[width=1\textwidth]{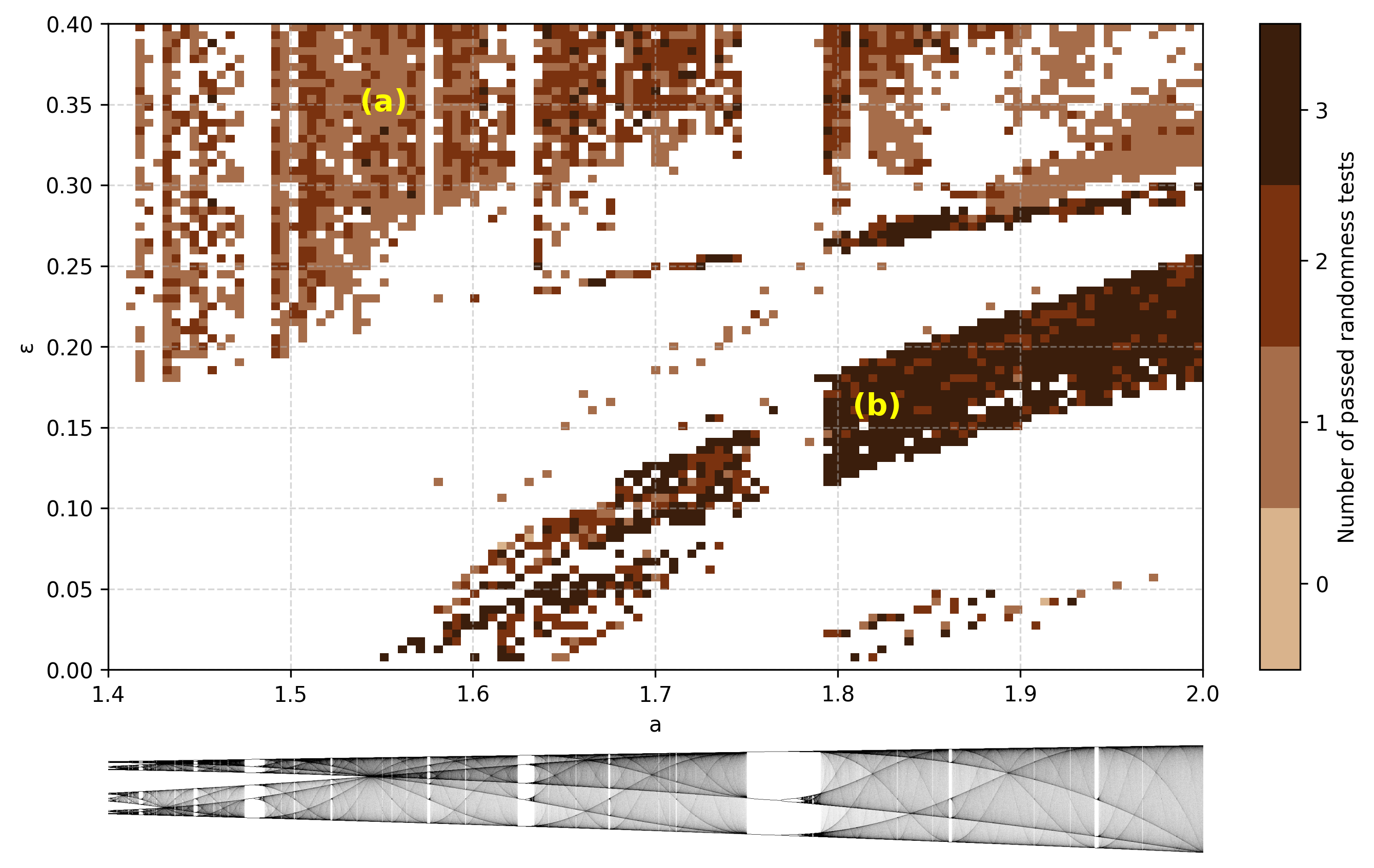}
\caption{Results of statistical tests for chaotic itinerancy in the GCM system with $N=3$ and a selection of different values of $a \in [1.4,2]$ and $\varepsilon \in [0.005,0.4]$. The bifurcation diagram of the logistic map is shown for reference. The regions (a) and (b) are discussed in the text.}
\label{stat_test}
\end{figure}

The next step in the search for chaotic itinerancy is the application of the randomness tests described in Section~\ref{sec:chaotic}, aimed at determining whether the transitions between the attractor ruins are truly chaotic or not. The number of tests passed by the sample trajectories for the parameters selected at the previous stage are shown in Figure~\ref{stat_test}. Each color-coded region indicates how many of the three statistical tests were passed for a given pair of parameters. Specifically, denote the three statistical tests \eqref{test:LB}, \eqref{test:ADF} and \eqref{test:OBD} by $S_1$, $S_2$ and $S_3$, respectively. For those parameters $a \in [1.4,2]$ and $\varepsilon \in [0.005,4]$ in the selected grid $121 \times 80$ that pass the first stage of verification of chaotic itinerancy, we compute the value
\[
\mathcal{N} (a,\varepsilon) = \mbox{card}\, \{i \in \{1,2,3\} : \text{the system passes the test $S_i$} \}.
\]
The values of $\mathcal{N} (a,\varepsilon)$ are shown in Figure~\ref{stat_test} with colors indicated in the color bar to the right of the plot, where the brightest shade corresponds to $\mathcal{N} (a,\varepsilon) = 0$ (none of the tests passed) and the darkest one to $\mathcal{N} (a,\varepsilon) = 3$ (all of the tests passed).

In region (b) in Figure~\ref{stat_test}, the results of the statistical tests confirm randomness in a majority of cases, especially to the right of the wide periodic window, whereas in region (a) the results are considerably worse. We checked that the difference was particularly due to failing the Ljung-Box test and the runs test. Although we do not show the results here, we remark that the computed variance of the local Shannon entropy and the variance of the local permutation entropy also reveal differences between the two regions. In region (b), these values are the highest in most cases, while in region (a) they are typically close to zero.

It turns out that there is a profound difference between the types of dynamics present in the system for the parameters in regions (a) and (b). The difference is in the dimension of the attractor ruins found in terms of high-density clusters in the phase space. The dimension of the attractor ruin reflects the type of synchronization between the one-dimensional maps that interact within the GCM system. We investigate this problem as follows.

We propose using Principal Component Analysis (PCA) \cite{jolliffe-pca} for the assessment of the dimensionality of the dynamics. We compute the fraction of the total variance captured by the first principal component obtained through PCA when analyzing the spatial distribution of all the points that form each attractor ruin. We describe this method in a general form applicable to a $p$-dimensional system.

Let $A$ be the set of points identified as one of the attractor ruins in the system with given parameter values $(a,\varepsilon)$. Consider the covariance matrix $\Sigma$ based on the points in $A$. This is a positive semi-definite matrix, so its eigenvalues are non-negative. Denote them by $\lambda_1 \geq \lambda_2 \geq \ldots \geq \lambda_p$. The relative variance captured by the first principal component computed from the set $A$ is defined as
\[
V_1(A) = \frac{\lambda_1}{\sum_{j=1}^{p} \lambda_j},
\]
which essentially means that projection of $A$ to the one-dimensional eigenspace spanned by an eigenvector corresponding to $\lambda_1$ captures $V_1(A)$ of the variability found in $A$. If $V_1(A)$ is large (e.g., over 95\%) then this means that the set $A$ is almost one-dimensional, that is, it is spanned closely to a line in the phase space.

\begin{figure}[htbp]
\centering
\includegraphics[width=1\textwidth]{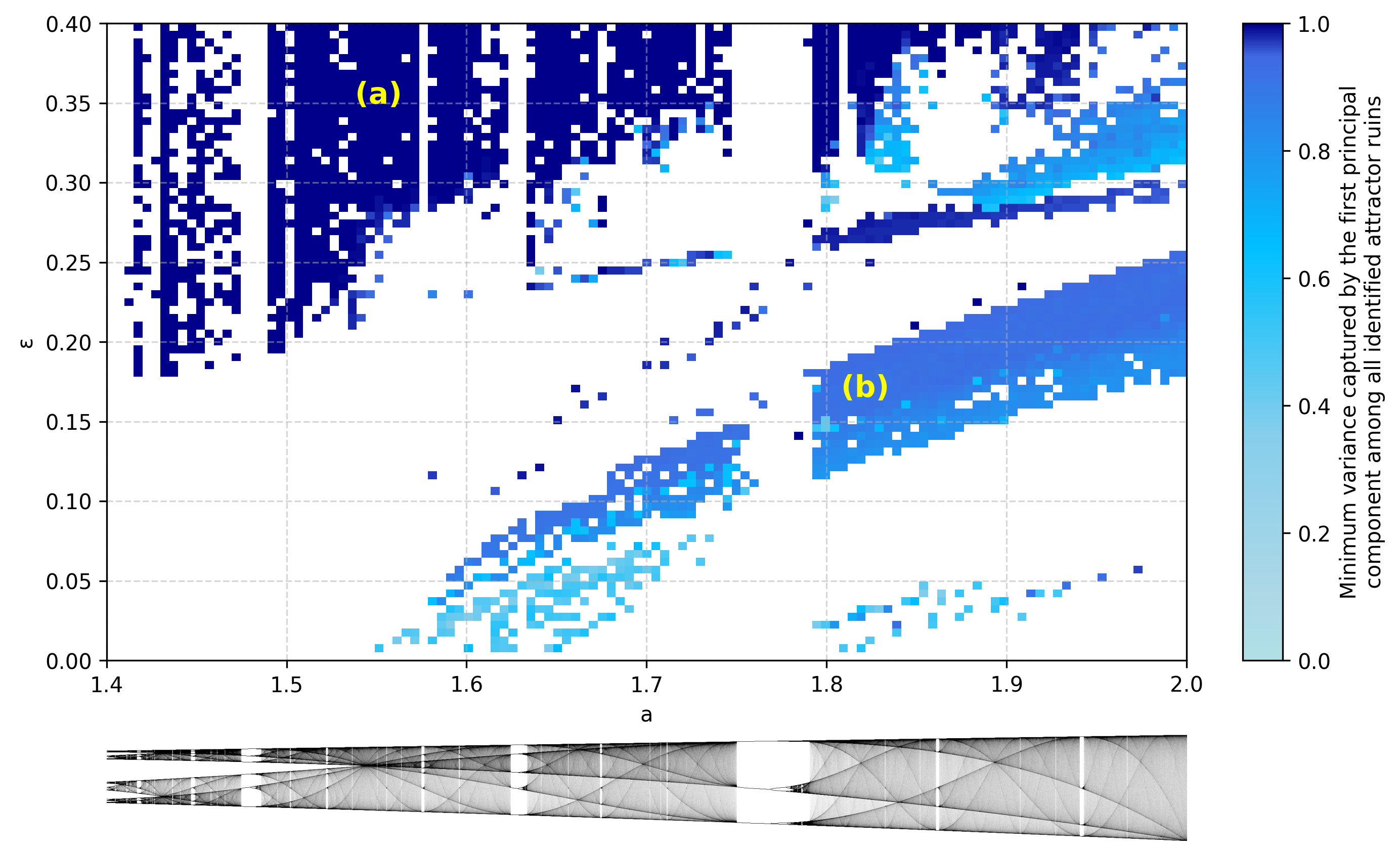}
\caption{Minimum fraction of the total variance captured by the first principal component in the coordinates of points among all the attractor ruins identified in the GCM system with $N=3$ and a selection of different values of $a \in [1.4,2]$ and $\varepsilon \in [0.005,0.4]$. Two specific regions are annotated in the diagram: (a) coherent phase and (b) intermittent phase.}
\label{pca}
\end{figure}

Figure \ref{pca} shows the minimum computed value $V_1(A)$ among all the identified attractor ruins for each pair of parameters $(a,\varepsilon)$ separately. Namely, for those parameters $a \in [1.4,2]$ and $\varepsilon \in [0.005,4]$ in the selected grid $121 \times 80$ that pass the first stage of verification of chaotic itinerancy, let us denote the set of all the attractor ruins found in the system with parameters $(a,\varepsilon)$ by $\mathcal{A}(a,\varepsilon)$, and compute the value
\[
\mathcal{D}(a,\varepsilon) = \min \{V_1(A) : A \in \mathcal{A}(a,\varepsilon)\}.
\]
The values of $\mathcal{D}(a,\varepsilon)$ are shown in the plot in Figure \ref{pca} using shades of blue, according to the color bar provided to the right of the plot. In particular, bright blue indicates lower values of $\mathcal{D}(a,\varepsilon)$, and dark blue---higher values of $\mathcal{D}(a,\varepsilon)$.

It turns out that region (a) in Figure \ref{pca} corresponds to the coherent phase in which all the coordinates are synchronized, as shown by the fact that nearly $100\%$ of the variability is covered by the first PCA component in all the attractor ruins found, and thus the identified attractor ruins are essentially one-dimensional. The opposite situation is encountered in region (b), where the dimension of at least one attractor ruin turns out to be larger than $1$, as judged by the PCA. This region corresponds to the intermittent phase located between the turbulent and coherent phases, where some attractor ruins have a higher dimension. The dimension higher than $1$ leaves enough room for the existence of complicated transitions between the attractor ruins, and distinguishes the dynamics from a one-dimensional chaotic system. Recall that the desired features of chaotic itinerancy in region (b) are also confirmed by the three statistical tests, as discussed earlier and shown in Figure~\ref{stat_test}.

Taking all this into account, we conclude that chaotic itinerancy is most evident in region (b) shown in Figures \ref{grid}--\ref{pca}, where the measures we propose indicate desirable chaotic behavior.

\section{Final remarks}
\label{sec:conclusion}

In this paper, we proposed a new methodology for investigating the phenomenon of chaotic itinerancy in semidynamical systems. We applied entropy-based techniques to identify parameter regimes likely to exhibit chaotic itinerancy. Then we used density-based clustering algorithms to find attractor ruins in specific systems. We demonstrated the effectiveness of this approach on a $3$-dimensional system of globally coupled logistic maps (GCM).

Although the phenomenon of chaotic itinerancy is often associated with high-dimensional systems, we were able to provide evidence for the presence of this phenomenon in the studied low-dimensional system. Indeed, some other low-dimensional systems are known in which chaotic itinerancy can be observed, like the two-dimensional model of mutually coupled Gaussian maps \cite{Kobayashi2017,Kobayashi2018-yv}. These results show that the phenomenon of chaotic itinerancy might appear in a multitude of dynamical systems, and therefore the development of methods for its detection is of wide interest.

It is remarkable that the results of the comprehensive analysis that we described in Section~\ref{sec:hdbscan} agree with the earlier observations made by Kaneko in 1990 \cite{Kaneko1990-fj}, but we obtained these results by applying a fully algorithmic procedure that did not require any visual inspection or heuristic assessment. We are therefore convinced that our method has considerable potential for practical analysis of a wide range of dynamical systems.

Finally, we would like to emphasize that our approach is in principle dimension-independent, so it is possible to use the proposed methodology to study high-dimensional systems. Definitions of local Shannon entropy and local permutation entropy are stated for systems in any dimension. However, if the number of coordinates is overwhelmingly high, one might compute these quantities restricted to a single coordinate or a small subset of coordinates in the hope that their variability reflects the overall behavior of the trajectories in the system. Moreover, density-based clustering is a concept that only requires a metric space, so it will work in an arbitrary dimension. However, the application of specific clustering algorithms in higher dimensions may be more computationally demanding. Nevertheless, with the immense increase in computing power of contemporary computers and the rapid development of machine learning techniques and algorithms, we are firmly convinced that our method is applicable to a growing class of dynamical systems of wide interest.

\section*{Author contributions (CRediT)}

\textbf{Nikodem Mierski:} Data curation, Formal analysis, Investigation, Methodology, Software, Validation, Visualization, Writing -- original draft, Writing -- review \& editing.
\textbf{Paweł Pilarczyk:} Conceptualization, Formal analysis, Funding acquisition, Investigation, Methodology, Project administration, Validation, Writing -- review \& editing.

\section*{Acknowledgments}

This research was supported by the National Science Centre, Poland, within the grant OPUS 2021/41/B/ST1/00405.

P.\ Pilarczyk expresses his gratitude to Professor Hiroshi Kokubu (Kyoto University) for introducing him to the topic of chaotic itinerancy.

\section*{Data availability statement}

All data shown and discussed in the paper can be generated by programs that have been made publicly available on~\cite{code}.

\printbibliography

\end{document}